\documentclass[longauth]{aa}
\usepackage[pdftex]{graphicx}
\usepackage{epsfig}
\usepackage{amsmath}
\usepackage{wasysym}
\usepackage{marvosym}
\usepackage{txfonts}
\pdfoutput=1
\usepackage[pdftex]{color,xcolor}
\usepackage[pdftex]{graphicx}
\usepackage[backref]{hyperref} 
\hypersetup{pdfauthor=Paul Molliere}
\hypersetup{backref=true, pagebackref=true, hyperindex=true, breaklinks=true,colorlinks=true,urlcolor=blue, linkcolor=blue,citecolor=blue,pagecolor=red, bookmarks=true, bookmarksopen=true}
\usepackage{epstopdf}
\hypersetup{backref=true}
\usepackage{lipsum}
\usepackage{natbib}
\usepackage{mathtools}
\let\bibfont\relax
\bibpunct{(}{)}{;}{a}{}{,} 
\usepackage{mhchem}

\def\ptrad{\emph{{petitRADTRANS}}}

\def\mearth{{\rm M}_\oplus}

\def\f1{f_{\rm I}}
\def\mj{{\rm M}_{\textrm{J}}}

\newcommand{\rj}{{\rm R}_{\rm J}}

\def\beq{\begin{equation}}
\def\eeq{\end{equation}}

\def\t2{\tau_{\rm II}}

\def\sigmas0{\Sigma_{\rm s,0}}

\def\petit{\emph{petitCODE}\ }

\def\apjl{ApJ}                
\def\apjs{ApJS}               
\def\ao{Appl.Optics}          
\def\aap{A\&A}                

\def\({\left(}
\def\){\right)}
\def\<{\left<}
\def\>{\right>}

\newcommand{\rch}[1]{#1}

\newcommand{\cend}{}

\begin{document}

\title{Retrieving scattering clouds and disequilibrium chemistry in the atmosphere of HR~8799e}

\author{P. Molli\`{e}re\inst{1,2}, T. Stolker\inst{3}, S. Lacour\inst{4,5,6}, G.~P.~P.~L. Otten\inst{7}, J. Shangguan\inst{5}, B. Charnay\inst{4}, T. Molyarova\inst{17}, M. Nowak\inst{8,9}, Th. Henning\inst{1}, G.-D. Marleau\inst{20,23,1}, D.~A. Semenov\inst{1,18}, E. van Dishoeck\inst{2,5}, F. Eisenhauer\inst{5}, P. Garcia\inst{10,13}, R. Garcia Lopez\inst{22,1}, J.~H. Girard\inst{16}, A.~Z. Greenbaum\inst{14}, S. Hinkley\inst{21}, P. Kervella\inst{4}, L. Kreidberg\inst{1}, A.-L. Maire\inst{11,1}, E. Nasedkin\inst{1}, L. Pueyo \inst{16}, I.~A.~G. Snellen\inst{2}, A. Vigan\inst{7}, {J. Wang\inst{12}}\thanks{51 Pegasi b Fellow}, P.~T. de Zeeuw\inst{2,5}, A. Zurlo\inst{15,19,7}}

\institute{Max-Planck-Institut f\"ur Astronomie, K\"onigstuhl 17, 69117 Heidelberg, Germany \and Leiden Observatory, Leiden University, Postbus 9513, 2300 RA Leiden, The Netherlands \and Institute for Particle Physics and Astrophysics, ETH Zurich, Wolfgang-Pauli-Strasse 27, 8093 Zurich, Switzerland \and LESIA, Observatoire de Paris, Universit\'e PSL, CNRS, Sorbonne Universit\'e, Univ. Paris Diderot, Sorbonne Paris Cité, 5 place Jules Janssen, 92195 Meudon, France \and Max Planck Institute for extraterrestrial Physics, Giessenbachstra{\ss}e 1, 85748 Garching, Germany \and European Southern Observatory, Karl-Schwarzschild-Stra{\ss}e 2, 85748 Garching, Germany \and Aix Marseille Univ, CNRS, CNES, LAM, Marseille, France \and Institute of Astronomy, University of Cambridge, Madingley Road, Cambridge CB3 0HA, United Kingdom \and Kavli Institute for Cosmology, University of Cambridge, Madingley Road, Cambridge CB3 0HA, United Kingdom \and CENTRA, Centro de Astrof\'{\i}sica e Gravita\c{c}\~{a}o, Instituto Superior T\'{e}cnico, Avenida Rovisco Pais 1, 1049 Lisboa, Portugal \and STAR Institute, Universit\'e de Li\`ege, All\'ee du Six Ao\^ut 19c, B-4000 Li\`ege, Belgium \and Department of Astronomy, California Institute of Technology, Pasadena, CA 91125, USA \and  Universidade do Porto, Faculdade de Engenharia, Rua Dr. Roberto Frias, 4200-465 Porto, Portugal \and Department of Astronomy, University of Michigan, Ann Arbor, MI 48109, USA \and N\'ucleo de Astronom\'ia, Facultad de Ingenier\'ia y Ciencias, Universidad Diego Portales, Av. Ejercito 441, Santiago, Chile \and Space Telescope Science Institute, Baltimore, MD 21218, USA \and Institute of Astronomy, Russian Academy of Sciences, 48 Pyatnitskaya St., Moscow, 119017, Russia \and Department of Chemistry, Ludwig-Maximilians-Universit\"at,
Butenandtstra{\ss}e 5-13, 81377 Munich, Germany \and Escuela de Ingenier\'ia Industrial, Facultad de Ingenier\'ia y Ciencias, Universidad Diego Portales, Av. Ejercito 441, Santiago, Chile \and  Institut f\"ur Astronomie und Astrophysik, Universit\"at T\"{u}bingen, Auf der Morgenstelle 10, 72076 T\"{u}bingen, Germany \and University of Exeter, Physics Building, Stocker Road, Exeter, EX4 4QL, United Kingdom \and School of Physics, University College Dublin, Belfield, Dublin 4, Ireland \and Center for Space and Habitability, Universit\"at Bern, Gesellschaftsstrasse 6, 3012 Bern, Switzerland}

\offprints{Paul MOLLIERE, \email{molliere@mpia.de}}

\date{Received -- / Accepted --}

\abstract
{Clouds are ubiquitous in exoplanet atmospheres and represent a challenge for the model interpretation of their spectra. Complex cloud models are too numerically costly for generating a large number of spectra, while more efficient models may be too strongly simplified.}
{We aim to constrain the atmospheric properties of the directly imaged planet HR 8799e with a free retrieval approach.}
{We use our radiative transfer code petitRADTRANS for generating spectra, which we couple to the PyMultiNest tool. We added the effect of multiple scattering which is important for treating clouds. Two cloud model parameterizations are tested: the first incorporates the mixing and settling of condensates, the second simply parameterizes the functional form of the opacity.}
{In mock retrievals, using an inadequate cloud model may result in atmospheres that are more isothermal and less cloudy than the input. Applying our framework on observations of HR 8799e made with the GPI, SPHERE and GRAVITY, we find a cloudy atmosphere governed by disequilibrium chemistry, confirming previous analyses. We retrieve that ${\rm C/O}=0.60_{-0.08}^{+0.07}$. Other models have not yet produced a well constrained C/O value for this planet. The retrieved C/O values of both cloud models are consistent, while leading to different atmospheric structures: cloudy, or more isothermal and less cloudy. Fitting the observations with the self-consistent Exo-REM model leads to comparable results, while not constraining C/O.}
{With data from the most sensitive instruments, retrieval analyses of directly imaged planets are possible. The inferred C/O ratio of HR 8799e is independent of the cloud model and thus appears to be a robust. This C/O is consistent with stellar, which could indicate that the HR 8799e formed outside the CO$_2$ or CO iceline. As it is the innermost planet of the system, this constraint could apply to all HR 8799 planets.}

\titlerunning{Scattering clouds and disequilibrium chemistry in the atmosphere of HR~8799e}
\authorrunning{P. Molli\`ere et al.}

\maketitle

\section{Introduction}
\label{sect:intro}
The description of clouds in exoplanets and brown dwarfs is one of the major uncertainties when modeling the structures and spectra of self-luminous atmospheres \citep[e.g.,][]{marley2013}. Fully modeling the microphysics of clouds is difficult, due to modeling uncertainties and long computational timescales, as the chemistry, nucleation process, condensation, particle coalescence, settling and mixing need to be accurately described. Moreover, even if all these processes are taken into account, it is not straightforward which values to prescribe for the remaining free parameters. Some of the ``free parameters'' of such elaborate cloud models are likely not free at all, but are determined by the full solution of the (multi-dimensional) atmospheric structure, which is a function of the cloud properties itself, due to the radiative feedback of the clouds. At the same time such complicated cloud models are useful and necessary, because they allow us to understand the interplay of physical processes during cloud formation, and will ultimately have to explain the cloud properties of all exoplanets and brown dwarfs, irradiated or self-luminous. Two examples for such complete cloud models are described in \citet{woitkehelling2004,hellingwoitke2008,woitkehelling2020} and \citet{gaomarley2018,powellzhang2018}.

When comparing synthetic cloudy spectra to observations, often one-dimensional self-consistent models are used, where the cloud opacity is radiatively coupled to the atmospheric temperature structure. Here iterating the structure is necessary. This requires a faster cloud modeling approach, which parametrizes parts of the cloud formation process. Examples are models based on timescale comparisons as in \citet{allard2001,allard2003}, implementing the approach of \citet{rossow1978}, or \citet{ackermanmarley2001}, which uses the ratio of the cloud particle settling and mixing velocities ($f_{\rm sed}$) as a free parameter, and solves for the particle size assuming a log-normal particle size distribution and a vertical diffusion coefficient $K_{\rm zz}$. The model of \citet{charnaybezard2018} is again different, and mixes the two previous approaches: as in \citet{ackermanmarley2001}, the vertical distribution of the cloud mass is determined assuming a steady state between mixing, settling and cloud condensation in every layer, while the average particle size is found using the timescale approach of \citet{rossow1978}. Thus no $f_{\rm sed}$ needs to be specified in \citet{charnaybezard2018} and this model can reproduce the L-T transition, including the effect of low gravity, which moves the transition to lower effective temperatures. Another interesting approach is the recent model by \citet{ormelmin2019}, which determines the cloud mass fraction and average particle size by solving the steady-state differential equations including cloud settling, mixing, nucleation, condensation and coagulation as a function of $K_{\rm zz}$ and the nucleation rate. A summary of other models can be found in \citet{hellingackerman2008,hellingcasewell2014}.

The two modeling philosophies described above (full microphysical or simplified for increased speed) are invaluable for understanding both cloud (micro)physics and the self-consistent radiative feedback of clouds. However, they are challenging when one aims at fitting cloudy spectra. The former, more complete models may take prohibitively long when calculating a large number of atmospheric structures and spectra. The latter, more parametrized models, allow for the calculation of larger self-consistent model grids. However, an important question is whether the simplification steps during model construction were all justified, and if all explicit (and implicit) free parameters of the model have been varied sufficiently. Moreover, an update of the model requires the calculation of a new grid, the models of which are demanding to produce and converge especially with clouds \citep[see, e.g.,][]{morleymarley2014,mollierevanboekel2016}, whereas considering additional parameters increases the grid size by orders of magnitude.

In the work presented here, we use a different approach. Namely we attempt to retrieve the characteristics of cloudy, self-luminous atmospheres by means of free retrievals. This is done by parameterizing the temperature profile, as well as the cloud properties, while using chemical equilibrium abundances with a simple quench pressure treatment to account for atmospheric mixing. While clouds in various parameterizations have been included in retrievals of transmission spectra of exoplanets \citep[see][for some recent examples]{mcdomadhu2017,fisherheng2018,tsiaraswaldmann2018,pinhasmadhusudhan2019,barstow2020}, cloudy retrievals are still a comparatively novel approach for self-luminous targets. The use of a free retrieval approach for fitting the spectra of brown dwarfs and directly imaged planets is motivated by \citet{lineteske2015,linemarley2017,zaleskyline2019}, who retrieved the atmospheric properties of clear T- and Y-dwarfs, \citet{burninghammarley2017}, who studied cloudy L-dwarfs, and \citet{leeheng2013,laviemendoca2017} who attempted to retrieve the properties of the cloudy HR~8799 planets for the first time. These pioneering works show the power of free retrievals for constraining condensation physics and clouds in cloud-free and cloudy brown dwarfs, and how retrieved planetary abundances may be connected to planet formation.

The radiative transfer tool used in the retrievals here is \ptrad \ \citep{mollierewardenier2019}, which we update to include the effect of scattering, which can no longer be ignored for cloudy atmospheres. By parametrizing the clouds, rather than making assumptions on how to simplify the cloud modelling process, we let the data constrain basic cloud characteristics such as cloud mass, location, and particle size, provided that the signal-to-noise of the data is high enough. Our model has the advantage that any changes in the (cloud) parametrization approach can be quickly implemented and tested, without the need of recalculating cloudy model grids.

The capabilities of our retrieval model are demonstrated by analyzing new and archival spectra of the cloudy planet HR~8799e, taken with the GRAVITY \citep[K band, see][]{lacournowak2019}, SPHERE \citep[YJH bands, see][]{zurlovigan2016}, and GPI \citep[H band, see][]{greenbaumpueyo2018} instruments. While not included in the fit, we also compare to archival mid-infrared (MIR) photometry.\footnote{\rch{Our rationale for excluding the photometry from the fit is explained in Section \ref{sect:descr_data}.}} The HR~8799 system is especially interesting because it hosts four directly imaged planets that orbit their star within a massive debris disk \citep{maroismacintosh2008,maroiszuckerman2010,currieburrows2011,surieke2009} at distances from 15 to 70~au \citep[e.g.,][]{wanggraham2018}. This allows for the comparative characterization of the planets' spectral properties. In particular the planets' atmospheric abundances may shed light on how they formed from the circumstellar disk.
Consequently, the HR~8799 planets have been extensively studied in the literature, and have been classified to bear the hallmarks of thick clouds and disequilibrium chemistry, placing them in the low-gravity, cool end of L spectral sequence (see Section \ref{sect:literature_comp} for a more detailed discussion of the literature). In addition to confirming these findings for HR~8799e, we also derive the planet's metallicity and, for the first time, carbon-to-oxygen number ratio (C/O), to study possible formation pathways.

Our retrieval model is detailed in Section \ref{sect:model}, including our description of the scattering implementation, the temperature and chemistry parameterization, as well as our cloud model parameterizations. Section \ref{sect:ver_ret} describes tests to verify our model setup with mock retrievals. Section \ref{sect:ret_hr8799e} contains our retrieval study of the planet HR8799e. Section \ref{sect:discussion} discusses the implications of our results for the formation of HR~8699e, and compares to the extensive body of literature on the HR~8799 planets. We summarize our findings in Section \ref{sect:summary} and provide an outlook for the further development and application of our new retrieval model.

\section{Forward retrieval model}
\label{sect:model}

\subsection{Adding scattering to \ptrad}
\label{sect:add_scat}

The objects we seek to model are expected to be inherently cloudy. Hence, scattering is an important process that needs to be considered during the radiative transfer calculations. To calculate quantities such as the photon destruction probability, it is necessary to compare the scattering and absorption opacities. For this, the total atmospheric opacity, combined from the individual absorber opacitites, needs to be calculated. Thus the correlated-k treatment of \ptrad \ needs to be adapted to combine the k-tables (opacity tables) of individual atmospheric absorbers.\footnote{Correlated-k means that the radiative transfer is carried out using the cumulative probability of the opacity distribution function as the spectral coordinate, and assuming that different probability values map to the same wavelength in all atmospheric layers \citep[e.g.,][]{lacis_oinas1991,fuliou1998,marleyrobinson2015}.} This is in contrast to the case treating purely emission, which can be handled by using the products of the transmissions of individual species \citep[see, e.g.,][]{irwinteanby2008,mollierewardenier2019}. Because our goal is to run retrievals, the k-table combination has to be as fast and as accurate as possible.

Here we present a newly developed method to quickly combine k-tables of different absorber species, which works by sampling the opacity distribution functions of individual absorbers. Computational time is saved by sampling the indices of the k-table entries, instead of interpolating the k-tables to sampled values of the cumulative probability. The k-table mixing process is described and tested in Appendix \ref{sect:ktab_mix}.

For solving the radiative transfer equation we then use the same treatment as described in our self-consistent \emph{petitCODE} \citep{mollierevanboekel2015,mollierevanboekel2016}, namely by using the Feautrier method \citep{feautrier1964}, converging the scattering source function with local Accelerated Lambda Iteration (ALI) \citep{olsonauer1986} and Ng acceleration \citep{ng1974}. The scattering process is assumed to be isotropic, with a $(1-g_{\rm a})$ correction factor applied to the scattering cross-sections, where $g_{\rm a}$ is the scattering anisotropy. The scattering implementation is further described in Appendix A.6 of \citet{mollierevanboekel2016}.
We show a verification of the cloudy spectra of \ptrad, including scattering, in Appendix \ref{sect:ver_scat}. For this verification, we used \petit to calculate a self-consistent HR~8799e model in radiative-convective and chemical equilibrium, which included clouds of \ce{MgSiO3} and \ce{Fe}. The spectra of \ptrad \ and \petit agree excellently. It takes \ptrad \ a few seconds to calculate a cloudy emission spectrum in the YJHK bands, which is fast enough for retrievals on computational clusters.

\subsection{Temperature model}
\label{sect:temperature_model}

Our goal is to parameterize the vertical temperature profile of the atmosphere in a way that imposes as few prior constraints on the solution as possible. It would hence appear to be ideal to retrieve the temperatures in every layer of the discretized atmosphere independently, which is an approach commonly followed in the planetary science community \citep[e.g.,][]{rodgers2000,irwinteanby2008}, and which has also been applied to cloud-free brown dwarfs \citep{linefortney2014} and exoplanets \citep[e.g.,][]{leefletcher2012}.

However, if the data are sparse or of low signal-to-noise, a level-by-level retrieval of the temperature can lead to overfitting and thus unphysical oscillations in the inferred temperature profile. One way of reducing such oscillations is by smoothing the resulting P-T profile \citep[as done in][]{irwinteanby2008,leefletcher2012}. Another way to circumvent this problem is to retrieve temperatures at a limited number of altitudes in the atmosphere, which are then connected via (spline) interpolation to yield a temperature at all layers, thereby reducing the number of free parameters \citep{lineteske2015,kitzmannheng2019}. In addition, \citet{lineteske2015} included a penalty term on the spatial sum of second derivatives of the temperature profile which further discouraged oscillatory solutions. Because the weight of said penalty can bias the results, its optimal value is also fitted in the inference process. This method has been used to retrieve the temperature profiles of cloud-free T- and Y-dwarfs in \citet{lineteske2015,linemarley2017,zaleskyline2019}.

The \rch{most biased} class of temperature models are those that use some kind of physical reasoning to parametrize the shape of the temperature profiles. This includes analytical solutions for self-luminous or irradiated atmospheres, assuming a gray or double-gray\footnote{That is, taking two (or more) separate gray opacities within given wavelength bands, for example in the optical and infrared.} opacity. The analytical solution (or a modification of it) by \citet{guillot2010,parmentierguillot2014} is commonly used, for example in \citet{linezhang2012,bennekeseager2012,lineknutson2013,linewolf2013,lineknutson2014,waldmannrocchetto2015,rocchettowaldmann2016,kreidbergline2018,brogiline2018,mollierewardenier2019}. There also exists the temperature parametrization suggested by \citet{madhusudhanseager2009}, which allows to parametrize temperature structures with or without inversions, and with or without a deep isothermal layer, as commonly expected for hot Jupiter planets. This parametrization has been used in, for example, \citet{madhusudhan2011,madhusudhanseager2011,madhusudhancrouzet2014,mcdomadhu2017,burninghammarley2017,gandhimadhusudhan2018,pinhasrackham2018,mcdomadhu2019}.

When investigating different models suitable for retrieving the atmospheres of cloudy self-luminous exoplanets, we settled on a model that uses both freely variable and physically motivated parameterizations, based on the atmospheric altitude. This temperature model, which allows us to retrieve the synthetic structures of cloudy atmospheres, is split into three parts, going from high, to middle, to low altitude. The spatial coordinate of the temperature model is an optical depth $\tau$\footnote{This optical depth is merely used for parameterization, so not associated to any particular wavelength or mean opacity.}, which we relate to the pressure $P$ by
\beq
\tau = \delta P^\alpha ,
\label{equ:tau_pt}
\eeq
where $\delta$ and $\alpha$ are free parameters, and $P$ is the atmospheric pressure in units of dyn cm$^{-2}$. This mapping is required because $P$ is the vertical coordinate of \emph{petitRADTRANS}. We then setup the atmospheric temperature profile, starting with the middle altitudes, that is, the `photosphere'.

\subsection*{`Photosphere' (middle altitudes)}
This region stretches from $\tau = 0.1$ to the radiative-convective boundary. Here we set the temperature according to the Eddington approximation
\beq
T(\tau)^4 = \frac{3}{4}T_{0}^4 \left(\frac{2}{3}+\tau\right) ,
\label{equ:t_edd}
\eeq
where $T_{0}$ is a free parameter. \rch{The optical depth $\tau$ is obtained from Equation \ref{equ:tau_pt} above.} In the original Eddington solution, from which we take the functional form of the temperature profile, this corresponds to the internal temperature. Likewise, we note that the `photospheric' region does not necessarily have to correspond to the true photosphere of the planet. Because the Eddington solution will always lead to an isothermal upper atmosphere, which is not expected to occur in reality, the high-altitude region of the atmosphere is treated separately, described immediately below.

\subsubsection*{High altitude}
This region extends from the top of the atmosphere ($P=10^{-6}$~bar) to $\tau = 0.1$. Here we split the atmosphere into four equidistant locations in ${\rm log}(P)$ space and treat the temperature at the three upper locations as free parameters. The temperature at the lowest altitude, which is at $\tau = 0.1$, is taken from the Eddington approximation of the `photosphere'. The temperatures in this atmospheric region are then found from a cubic spline interpolation.

\subsection*{Troposphere (low altitudes)}
This region starts from the radiative-convective boundary and extends to the bottom of the atmosphere. The radiative convective boundary is found by comparing the atmospheric temperature gradient of the Eddington approximation with the moist adiabatic temperature gradient of the atmosphere. This is done by interpolating the moist adiabatic gradient in the $T$-$P$-[Fe/H]-C/O space of the chemistry table (see Section \ref{sect:chem_model}). As soon as the atmosphere is found to be Schwarzschild-unstable, the atmosphere is forced onto the moist adiabat.

\subsection*{Priors}
\label{sect:priors}
We restrict $\alpha$ (see Equation \ref{equ:tau_pt}) to vary between 1 and 2, following \citet{robinsoncatling2012}. Moreover, to prevent the formation of temperature inversions, which are not expected in self-luminous objects, we required that the three free temperature points in the high altitude region of the atmosphere are colder than the highest point of the `photosphere' (middle altitude region), and that they decrease in temperature monotonically with increasing altitude. This prior is enforced by setting the upper boundary of the allowed temperature range of such a free temperature point equal to the temperature of the underlying temperature point, identical to the treatment in \citet{kitzmannheng2019}.

In \citet{nowaklacour2020} we had found that we had to impose further priors on the temperature parameterization, based on the structure of the atmospheric opacity of a given forward modeling realization. For example, the power law index of the optical depth, $\alpha$, was not allowed to deviate too far from the power law index measured from the opacity structure in the forward model, within the spectral range of the retrieved data. We found that this was necessary because of the high dimensionality of our retrieval model, and our inability to make the MCMC sampler find the global maximum of the log-probability in a finite amount of time otherwise. These opacity priors restricted the parameter space for the MCMC sufficiently. In this publication we use nested sampling \citep{skilling2004}, and using a sufficiently large number of live points made such opacity priors unnecessary.

\subsection{Chemistry model}
\label{sect:chem_model}

In the retrievals presented in this work, the chemical abundances within the atmosphere are determined by means of interpolation in a chemical equilibrium table, with a simple quench layer approximation used to account for atmospheric mixing. The abundance tables are  calculated with \emph{easyCHEM}, our \emph{CEA} \citep{gordon1994,mcbride1996} clone described in \citet{mollierevanboekel2016}.  The equilibrium condensation of the following species is included in the abundance calculations: \ce{Al2O3}, \ce{Fe}, \ce{FeO}, \ce{Fe2O3}, \ce{Fe2SiO4}, \ce{H2O}, \ce{H3PO4}, \ce{KCl}, \ce{MgSiO3}, \ce{Mg2SiO4}, \ce{Na2S}, \ce{SiC}, \ce{TiO}, \ce{TiO2}, \ce{VO}. Because rainout is expected to remove Si from the upper layers of the atmosphere \citep[see, e.g.][]{lodders2010}, we do not include feldspars, thereby inhibiting the sequestration of Na and K at high temperatures. This is consistent with abundance constraints of the alkalis from systematic retrieval analyses of T- and Y-dwarfs \citep{linemarley2017,zaleskyline2019}.

The chemical abundances (mass fractions) are tabulated as a function of pressure $P$, temperature $T$, carbon-to-oxygen number ratio C/O, and metallicity [Fe/H]. The pressure ranges from $10^{-8}$ to 1000 bar, in 100 points spaced equidistantly in ${\rm log}(P)$ space. The temperature ranges from 60 to 4000 K, in 100 equidistant points. The C/O values go from 0.1 to 1.6, in 20 equidistant points and the metallicity is tabulated for [Fe/H] values going from -2 to 1.84, in 31 equidistant points. Four-dimensional linear interpolation is used to interpolate the log-abundances of all absorbers. If a $T$-$P$-[Fe/H]-C/O coordinate falls outside of the grid, the abundances interpolated to the closest boundary point are used. The C/O ratio is varied by varying the oxygen abundance. For minimizing the Gibbs free energy we use the thermodynamic data of the \emph{CEA} code, or the references detailed in \citet{mollierevanboekel2016}. The data for FeH were obtained from M. Line (priv. comm.).

We approximate the effect of disequilibrium chemistry by setting the quench pressure $P_{\rm quench}$ as a free parameter. For atmospheric pressures $P<P_{\rm quench}$ we take the abundances of \ce{CO}, \ce{H2O}, and \ce{CH4} to be constant, and equal to the abundances at $P=P_{\rm quench}$. This follows the result from, for example, \citet{zahnlemarley2014}, namely that the abundances of \ce{CO}, \ce{H2O}, and \ce{CH4} can be taken to be constant above the quenching point. This treatment has been further verified by comparing to the results of the reaction network by \citet{venot2012,venot2015} in \citet{baudinomolliere2017}. 

The chemical abundance table also contains the value of the adiabatic temperature gradient, $\nabla_{\rm ad}$, which was calculated with \emph{easyCHEM}, using Equations 2.50, 2.59, and 2.75 of \citet{gordon1994}. The derivatives used for the calculation of the specific heat of the mixture are so-called equilibrium derivatives \citep{gordon1994}, meaning that the $\nabla_{\rm ad}$ value used in this work accounts for any change in the abundances (and thus heat release) during the adiabatic temperature change. Hence our adiabats are moist adiabats. The interpolation in the $\nabla_{\rm ad}$ table is used when constructing the temperature profile in the troposphere of our temperature model.

\subsection{Cloud model 1}
\label{sect:cloud_model_1}

As discussed in Section \ref{sect:intro}, our goal is to impose as few prior assumptions on the cloud properties as possible. Hence the ideal setup would be to freely retrieve the altitude-dependent distribution of cloud particle radii, as well as the vertical cloud density for every layer, independently. Such an approach would require many free parameters, which in turn requires enough data points of sufficiently high signal-to-noise to prevent over-fitting. Instead, we start with a more modest approach. As the \citet{ackermanmarley2001} model is already implemented in \ptrad, we use its three free parameters to control the mean particle size, cloud mass fraction and particle size distribution independently.

Hence, for the cloud to be retrieved, we set the settling parameter $f_{\rm sed}$ as a free parameter, which controls the altitude-dependent cloud mass fraction $X^{\rm c}$ above the cloud base via
\beq
X^{\rm c}(P) = X^{\rm c}_0 \left(\frac{P}{P_{\rm base}}\right)^{f_{\rm sed}} \ ,
\eeq
where the cloud mass fraction at the cloud base $X^{\rm c}_0$ is an additional free parameter. The pressure at the cloud base $P_{\rm base}$ is found by intersecting the saturation vapor pressure curve of the considered cloud species with the temperature profile of the atmosphere. \rch{In our retrievals below, we only consider \ce{MgSiO3} and \ce{Fe} clouds, because silicates and iron likely dominate the atmosphere at the temperatures and surface gravities reported for HR~8799 planets (see, e.g., Figure 7 in \citealt{morleyfortney2012}, and \citealt{marleysaumon2012,charnaybezard2018}). The recent findings of \citet{gaothorngren2020}, based on micro-physical cloud modeling, corroborate the importance of especially silicate clouds at these temperatures.}

We also let the vertical eddy diffusion coefficient $K_{zz}$ vary as a free parameter, which effectively sets the particle size, given an $f_{\rm sed}$ value. In self-consistent calculations, the $K_{\rm zz}$ parameter is usually set by mixing length theory with some lower limit assumption \citep[see, e.g.,][]{ackermanmarley2001,morleymarley2014,mollierevanboekel2016,charnaybezard2018}, or fixed to a constant value \citep[e.g.,][]{marleysaumon2012,samlandmolliere2017}. Here we let it float as a free parameter (and take it to be vertically constant), so as to determine the average particle size independently from $f_{\rm sed}$. We note that a retrieved $K_{\rm zz}$ value could be inconsistent with the derived chemical quench pressure (see Section \ref{sect:chem_model}). This could imply either a true shortcoming of our retrieval model, a shortcoming in a disequilibrium kinetic network, or a deviation from how $K_{\rm zz}$ actually sets the average particle size to how it is implemented in the \citet{ackermanmarley2001} model.

Lastly, the particle size distribution is fitted by letting the width of the log-normal size distribution, $\sigma_{\rm g}$, be a free parameter. This parameter is usually not varied if the \citet{ackermanmarley2001} cloud model is used. We note that it has been shown that a log-normal particle size distribution can be a poor choice when compared to the often bi-modal particle size distributions found from microphysics \citep{gaomarley2018,powellzhang2018}. However, in \citet{gaomarley2018}, the $K_{zz}$ and $\sigma_{\rm g}$ values were both fixed when fitting $f_{\rm sed}$ to the cloud structure of the microphysics result. Here we let $f_{\rm sed}$, $K_{\rm zz}$, and $\sigma_{\rm g}$ vary independently, so the retrieval should be flexible enough to determine the values of these three parameters that describe the cloud mass fraction, effective particle size, and dispersion of sizes around that value, independently. In principle, this model can also describe mono-disperse particle distributions, if a retrieval were to favor cases with $\sigma_g$ close to 1. In the limit $\sigma_g \rightarrow 1$ a log-normal particle size distribution approaches a delta function. In general, the retrieved cloud parameter values are expected to describe those visible atmospheric layers which are most affected by the clouds.

\subsection{Cloud model 2}
\label{sect:cloud_model_2}
The choice of $f_{\rm sed}$, $K_{zz}$ and $\sigma_{\rm g}$ in Cloud Model 1 may just be a glorified way of setting the cloud spectral slope and single scattering albedo. It is also questionable, for example, whether a retrieved $K_{zz}$ does actually correspond to the true vertical diffusion coefficient of the atmosphere. Rather, it could effectively be a nuisance parameter of the retrieval, varied to mimic the true properties of the cloud opacity, as alluded to above.

To test for a less physically motivated treatment, we constructed Cloud model 2. This approach is motivated by the cloud parameterization of \citet{burninghammarley2017}. Our model retrieves the spectral slope $\xi$ (which we take to be vertically constant) of the cloud opacity directly. Specifically, we set
\beq
\kappa_{\rm tot} = \kappa(P) \left(\frac{\lambda}{\rm 1 \ \mu m}\right)^\xi ,
\label{equ:spec_slope}
\eeq
where $\kappa_{\rm tot}$ is the total (scattering + absorption) cloud opacity, $\kappa(P)$ is its value at 1~$\mu$m, at pressure $P$, and $\lambda$ the wavelength.
The $\kappa(P)$ we describe as
\beq
\kappa(P) = \kappa_0 \left(\frac{P}{P_{\rm base}}\right)^{f_{\rm sed}} \ {\rm for} \ P <  P_{\rm base}, \label{equ:degen_kappa_P}
\eeq
and set it to zero for pressures larger than the cloud base pressure $P_{\rm base}$. The $f_{\rm sed}$ again describes the power law decrease of the cloud with altitude. Additionally, we set the single-scattering albedo, {$\omega$}, as a free parameter, which we also take to be vertically constant. For the absorption opacity it then holds that
\beq
\kappa_{\rm abs} = (1-{\omega})\kappa_{\rm tot}.
\eeq

In summary, the five free parameters of Cloud model 2 are $\kappa_0$, $\xi$, $f_{\rm sed}$, $P_{\rm base}$, and ${\omega}$.
Because there is a degeneracy between $\kappa_0$ and $P_{\rm base}$ (see Equation \ref{equ:degen_kappa_P}) if the atmosphere below the cloud deck cannot be probed by the observations, we put a prior on $P_{\rm base}$ such that
\begin{multline}
{\rm log}(L_{\rm base}) =  {\rm log}\left[{\rm exp}\left(-\frac{{\rm log}_{10}^2(P_{\rm base}/P_{\rm Fe})}{2\cdot (0.5 \ {\rm dex})^2}\right)+\right. \\
\left.{\rm exp}\left(-\frac{{\rm log}_{10}^2(P_{\rm base}/P_{\rm MgSiO_3})}{2\cdot (0.5 \ {\rm dex})^2}\right)\right] ,
\label{equ:pbase}
\end{multline}
where $P_{\rm Fe}$ and $P_{\rm MgSiO_3}$  are the cloud base positions of \ce{Fe} and \ce{MgSiO3}, respectively, obtained from intersecting their saturation vapor pressure curves with the atmospheric temperature profile.
Clearly, this will also tend to favor clouds that lie close to the expected cloud base positions for these condensate species. \rch{We note that this prior choice was made to not introduce parameters which are degenerate by construction. As mentioned, silicate and iron clouds may be the dominant cloud opacity carriers for the temperatures and surface gravities generally inferred for the HR~8799 planets. Removing this prior would allow to better probe situations where different cloud species dominate in the planetary atmosphere. This  will be tested in our future work.}

To test how well such a cloud description may be suited to describe more ``physically consistent'' clouds we carried out the following test. The cloud properties of the atmospheric model we used to create the synthetic observation for the verification retrieval (see Section \ref{sect:ver_ret}) were generated with Cloud model 1. Fitting these cloud opacities with Equation \ref{equ:degen_kappa_P}, and the spectral slope with Equation \ref{equ:spec_slope}, we found that the cloud parameters could indeed be well represented with Cloud model 2. In addition, the single-scattering albedo, taking the spectral average over the 0.9 to 2.5~$\mu$m range, was consistent with a large (${\rm \omega}\sim 0.85$), vertically constant value.
\begin{table}[t!]
\centering
{ \footnotesize
\begin{tabular}{ll|ll}
\hline \hline
Parameter & Value & Parameter & Value \\ \hline
$T_1$ & 330.6 K & ${\rm log}(X^{\rm Fe}_0/X^{\rm Fe}_{\rm eq})$ & -0.86 \\
$T_2$ & 484.7 K & ${\rm log}(X^{\rm MgSiO_3}_0/X^{\rm MgSiO_3}_{\rm eq})$ & -0.65\\
$T_3$ & 687.6 K & $f_{\rm sed}$ & 3\\
${\rm log}(\delta)$ & -7.51 & ${\rm log}(K_{\rm zz}/{\rm cm^2 s^{-1}})$ & 8.5 \\
$\alpha$ & 1.39 & $\sigma_{\rm g}$ &  2\\
$T_0$ & 1063.6 K & $R_{\rm P}$ & 1~$\rj$ \\
$\rm C/O$ & 0.55 & ${\rm log}(g/{\rm cm \ s^{-2}})$ & 3.75 \\
$\rm [Fe/H]$ & 0 & ${\rm log}(P_{\rm quench}/{\rm bar})$ &  -10 \\ \hline
\end{tabular}
}
\caption{Parameters for generating the synthetic observations of the cloudy exoplanet spectrum  retrieved for verification purposes in Section \ref{sect:ver_ret}. $X_{\rm eq}$ is the mass fraction predicted for the cloud species when assuming equilibrium condensation at the cloud base location.}
\label{tab:ret_input}
\end{table}

\section{Verification}
\label{sect:ver_ret}
\subsection{Retrieval with Cloud model 1}
\label{sect:cloud_model1_test_retrieval}
Here we present our retrieval tests when using Cloud model 1 (see Section \ref{sect:cloud_model_1}), that is, \ce{MgSiO3} and Fe clouds parameterized using the \citet{ackermanmarley2001} model, varying all of its three free parameters. We generated a synthetic observation as follows: for the input parameters of the temperature profile, we fitted our temperature model to the self-consistent atmospheric structure used for verifying our scattering implementation (see Section \ref{sect:add_scat} and Appendix \ref{sect:ver_scat}). The values of all input parameters of the model are shown in Table \ref{tab:ret_input}.

\begin{figure*}[t!]
\centering
\includegraphics[width=0.78\textwidth]{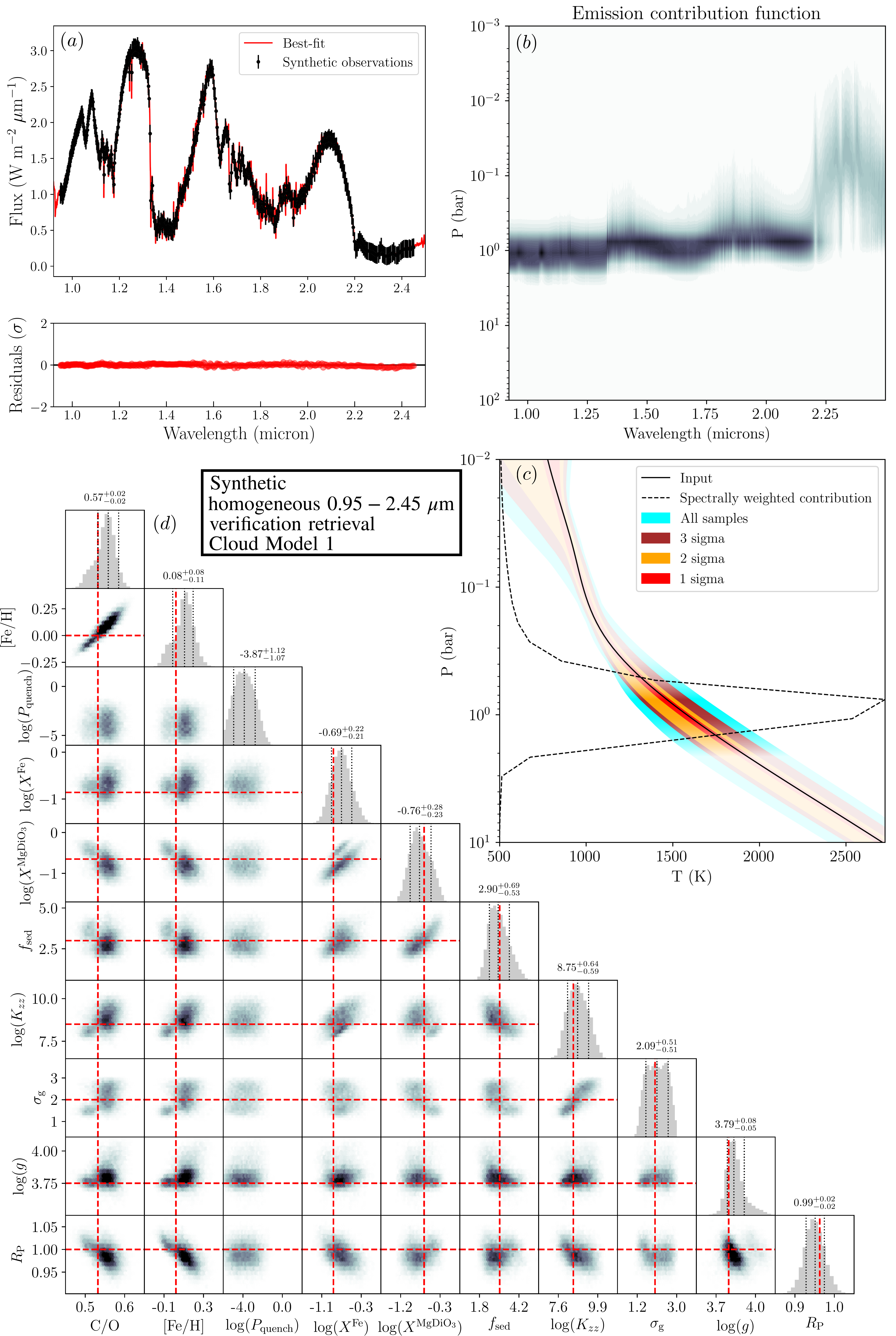}
\caption{Results of the verification retrieval using Cloud model 1. {\it Panel (a):} synthetic observation, best-fit spectrum and residuals. {\it Panel (b):} emission contribution function. Due to the clouds, pressures larger than 1-2 bar cannot be probed. {\it Panel (c):} retrieved pressure-temperature confidence envelopes. The black dashed line shows the flux average of the emission contribution function that is shown in Panel (b). The opaqueness of the temperature uncertainty envelopes has been scaled by this contribution function, with a minimum value of 10~$\%$. {\it Panel (d):} 2-d posterior plot of the (non-nuisance) retrieved atmospheric parameters. The red dashed lines denote the input values. The values of the cloud mass fractions at the cloud base have been divided by the mass fractions predicted when assuming equilibrium condensation at the cloud base location.}
\label{fig:combine_all_alpha_12_1000_noise_free}
\end{figure*}

To derive the posterior abundances of our fit we used the PyMultiNest\footnote{\url{https://johannesbuchner.github.io/PyMultiNest/}} package \citep{buchnergeorgakakis2014}, which is a Python wrapper of the MultiNest method \citep{ferrozhobson2008,ferrozhobson2009,ferrozhobson2013} for nested sampling \citep{skilling2004}. Nested sampling has the benefit of being able to approximate model evidences (i.e., the probability of the model, given the data), which allows for the pair-wise vetting of different models. Moreover, it is sampling the parameter space more thoroughly. This minimizes the problem of sampling the posterior distribution around a local, but not the global, maximum of the log-probability. It does not fully alleviate this problem, however (see discussion below). To ensure a high sample acceptance fraction of our high-dimensional model, we ran MultiNest in the constant efficiency mode, with a prescribed sampling efficiency of 5~\%. When using MultiNest in Importance Nested Sampling mode, evidences can still be calculated, even when prescribing a sampling efficiency\footnote{Also see the corresponding discussion in the MultiNest manual at \url{https://github.com/farhanferoz/MultiNest/blob/master/README.md}.}. We used 4000 live points in our retrievals, which we found necessary to cover the parameter space sufficiently.

To initially test our retrieval framework under idealized conditions, a synthetic observation was created assuming a continuous wavelength spacing of $\lambda/\Delta \lambda=400$ between 0.95 and 2.45~$\mu$m. We focused on this spectral region because it overlaps with the YJHK-bands of the SPHERE, GPI and GRAVITY instruments. The flux error was chosen to be constant across this wavelength range, with a mean S/N value of 10 per wavelength step. For comparison, the S/N per wavelength step of HR~8799e is about 4 ($\lambda/\Delta \lambda\approx 70$), 7 ($\lambda/\Delta \lambda\approx 200$), and 11 ($\lambda/\Delta \lambda\approx 1000$) for the SPHERE YJH \citep{zurlovigan2016}, GPI H \citep{greenbaumpueyo2018} and GRAVITY K \citep{lacournowak2019} band data, respectively. In order to not be affected by a given noise instantiation during the verification retrieval, we took the observational errors into account for calculating the log-likelihood, but did not perturb the mock observations using these error bars.

The results of this verification retrieval are shown in Figure \ref{fig:combine_all_alpha_12_1000_noise_free}. Panel (a) shows the synthetic observation, best-fit model, and the residuals between the two, scaled by the error bars. The residuals are flat and consistent with zero.

The emission contribution function of the best-fit model is shown in Panel (b) of Figure \ref{fig:combine_all_alpha_12_1000_noise_free}. Regions between 0.004 and 2 bar are accessible, with the \ce{Fe} and \ce{MgSiO3} cloud blocking the flux from the deeper regions. Methane absorption blocks most of the flux longward of 2.2~$\mu$m, probing cool regions as high as 0.004 bar. Shortward of 2.2~$\mu$m most of the flux originates from a narrow pressure region from 0.2 to 2 bar.

In Panel (c) the retrieved pressure temperature structure of the atmosphere is shown, with the percentiles setting the boundaries of the uncertainty envelopes corresponding to the 1-, 2- and 3-$\sigma$ ranges of a Gaussian distribution. In order to illustrate which altitudes of the atmosphere can actually be probed by the observation, the opaqueness of the temperature uncertainty envelopes has been scaled by the atmospheric contribution function, with a minimum value of 10~$\%$. For this the contribution function was flux-averaged. As can be seen, the uncertainty envelopes follow the input P-T profile. The input profile lies within the 1-$\sigma$ envelope.

Finally, Panel (d) shows the corner plot of the remaining parameters. The planetary radius, surface gravity, metallicity and C/O ratio can all be well retrieved. Only an upper limit is found for the quench pressure, which is as expected, as no quenching was considered. The cloud parameters are well constrained. We note that the retrieval appears to have found a bi-modal parameter distribution, but the one-dimensional posteriors constrain the input parameters well. A clear positive correlation can be seen between the metallicity and C/O ratio. This is attributed to the fact that we vary the oxygen abundance when changing C/O, such that a higher C/O corresponds to less oxygen, and hence water.

In summary, the retrieval verification test presented here can be regarded as successful. We were able to retrieve the temperature, composition, and cloud properties of the atmosphere. Nonetheless, the following challenges can be identified: even though an excellent fit to the spectrum has been achieved (see Panel (a) of Figure \ref{fig:combine_all_alpha_12_1000_noise_free}) the median values of the retrieved parameters were not exactly at the input values (although within the 1~$\sigma$ envelope). This is unexpected because the synthetic observation was not perturbed by the assumed error bars, in order to not be sensitive to stochastic noise of a given noise instantiation. Another point is the bi-modality of the posterior.

In our initial tests we found that using 400 live points in PyMultiNest resulted in biased retrieval results, such that median parameter values could be more than 1~$\sigma$ away from the input values, and residuals in the spectral fit that were larger. Increasing the number of live points to 4000 ameliorated this issue: the input parameters were retrieved at higher accuracy (within 1$\sigma$), and the residuals to the best-fit spectrum became smaller. We deduce from this that for input models of high dimensionality a sufficient number of live points has to be used, so as to increase chances that the positions of the live points sampled during the early stages of the nested sampling run will fall into the vicinity of the global maximum of the likelihood. Because the nested sampling method will zero-in on the highest likelihood regions during the retrieval, the danger exists that the true maximum location in parameter space will be missed. This problem is especially pressing for observations of high S/N, such as used in our example presented here, because the high-likelihood volume of the parameter space will shrink. This means that higher quality observations require a larger number of live points, and thus more computational time. This is especially important for the large spectral coverage, high S/N data to be taken with JWST.

Further exploring the bi-modality of the posterior shown in Figure \ref{fig:combine_all_alpha_12_1000_noise_free}, we ran a second retrieval with smaller prior ranges. They were restricted by the high-likelihood regions of the posterior from the initial fit, enclosing its bi-modal posterior distribution. The resulting posterior is shown in the corner plot in Figure \ref{fig:corner_tight_box}. It is unimodal and consistent with the input parameters.
In general, the bimodality and the offset between the median and input parameters of the posteriors indicate that multiple parameter combinations can lead to excellent spectral fits, while within 1~$\sigma$ of the true parameters. This may indicate that a retrieval model with fewer free parameters could be favored, leading to unique solutions, which we will explore in future studies.
\cend

\subsection{Retrieving Cloud Model 1 with Cloud model 2}
\label{sect:ret_CM1_with_CM2}

In this section we describe what happens when retrieving a mock observation made with Cloud Model 1 using Cloud Model 2. The retrieval was thus set up identically to the one described in the section immediately above, but used Cloud Model 2, while the mock observation was identical to the test above, that is made with Cloud Model 1. Figure \ref{fig:combine_all_alpha_12_simple_cloud} shows the corresponding results (see Section \ref{sect:ret_ver_test_cloud_model_2}).

We find that the spectral fit is again very good, although there are a few regions of systematic residuals of 1~$\sigma$. Hence, Cloud model 2 appears to satisfactorily describe the properties of the synthetic observations generated with Cloud model 1. However, we find significant differences in the retrieved atmospheric properties. C/O is constrained to $0.58_{-0.01}^{+0.01}$ (input was 0.55), [Fe/H] is retrieved to be $0.14_{-0.08}^{+0.08}$ (input was 0), and ${\rm log}(g)=4.01_{-0.10}^{+0.10}$ (input was 3.75). Thus, we find values close to, but offset from the true input values.

Moreover, instead of probing down to 2~bar at most, the atmosphere can now be probed down to 10 bar, and is more isothermal than the input temperature profile: here the retrieval mimics the effect of a thick cloud. Instead of the cloud hiding the deep hot regions from view, these regions are erroneously constrained to be less hot by the retrieval. At the same time the cloud is thus too optically thin and deep, with the retrieved cloud position at ~8 bar (which is well constrained). The emission contribution consequently shows that the emission stems from a more extended region than in the retrieval described in the section immediately above. Hence Cloud Model 2 was not able to describe the clouds made with Cloud Model 1 accurately enough, such that the retrieval modified the $P$-$T$ profile instead.

We conclude that a good fit to the spectrum alone is a dangerous measure when assessing whether or not a fit result is reasonable. All retrieved parameter values need to be carefully vetted, the retrieved P-T profile should also be compared to that of a self-consistent atmospheric code, when running the latter using the best-fit or median parameters of the free retrieval. We note that with free retrievals alone it may also be challenging to determine whether the atmospheric temperature profile is truly shallower than expected, while being less cloudy at the same time. This has been suggested by \citet{tremblinamundsen2015,tremblinamundsen2016,tremblinchabrier2017}, challenged in \citet{leconte2018}, and defended in \citet{tremblinpadioleau2019}. As they suggested, we find here that a shallow $P$-$T$ profile can indeed result in an excellent fit to the observations, even though we know that in this case the input model was actually more cloudy than retrieved. At the same time it is somewhat reassuring that in our example shown here the absolute deviation between the input and retrieved parameters such as C/O, [Fe/H], ${\rm log}(g)$, etc. is not large. The values, however, are biased, and we stress that a more detailed study needs to be performed on how strongly cloud model assumptions can affect the retrieved best-fit parameters.

\section{Retrieving HR~8799e}
\label{sect:ret_hr8799e}
In this section we describe how we used GRAVITY and archival SPHERE and GPI data to retrieve the atmospheric properties of HR~8799e, which is located at 15~au from its host star \citep[e.g.,][]{wanggraham2018}. We also compare our results to archival photometry for this planet, while not including these data in the retrieval itself.

\subsection{Data}
\label{sect:descr_data}

\subsubsection*{GRAVITY K band spectroscopy}

\begin{table}[t!]
\centering
{ \footnotesize
\begin{tabular}{llcrrr}
\hline 
Date & target & Exp & NDIT & DIT & seeing\\ 
\hline \hline
2018 Aug 28 & HR\,8799\,e & 7 & 100 & 10\,s & 0.5-0.8"\\ 
$\cdot$ & HR\,8799\,A & 2 & 50 & 1\,s \\ 
\hline
2019 Nov 9 & HR\,8799\,e & 3 & 60 & 8\,s & 0.8-1.0"\\ 
$\cdot$ & HR\,8799\,A & 3 & 64 & 1\,s \\ 
\hline
2019 Nov 11 & HR\,8799\,e & 3 & 100 & 8\,s & 0.8-1.1"\\ 
$\cdot$ & HR\,8799\,A & 1 & 64 & 1\,s \\ 
$\cdot$ & HD\,25535\,AB & 8 & 64 & 1\,s \\
\hline
\end{tabular}
}
\caption{Log of the GRAVITY observations.}
\label{tab:logGRAVITY}
\end{table}

We use three separate GRAVITY \citep{2017A&A...602A..94G} observations of HR8799e. First, the observation presented in \citet{lacournowak2019}. In addition, we here report on two new observations, taken November 9th, 2019 and November 11th. 2019, as part of the ExoGRAVITY Large Program. The log of the observations is presented in Table\ \ref{tab:logGRAVITY}. The observations on the 9th were made from a short observing block, with a total integration time of 180\,s. The phase referencing was done on the star using the fringe tracker \citep{2019A&A...624A..99L}. The zero point of the metrology was obtained by directly observing the star on the spectrometer. The observations on the 11th were done using the roof mirror as a field splitter: 100\% of the planetary flux could be used, but such an observation needs a binary to calibrate the zero point. This zero point was obtained on the binary system HD\,25335.

The data were reduced analogously to the data reduction presented for $\beta$~Pic~b in \citet{nowaklacour2020}. The flux of stellar origin is removed, and the spectra were obtained from the ratio between the coherent flux on the planet from the coherent flux on the star. This ratio is then multiplied by a theoretical BT-NextGen spectra of the star \citep{2012RSPTA.370.2765A}. The spectra is therefore calibrated from the telluric absorption. For the retrieval, the full spectral covariance is considered when deriving log-likelihoods.

\subsubsection*{SPHERE and GPI archival data}
We use the YJH-band spectroscopy of SPHERE reported in \citet{zurlovigan2016}. In addition, we use the GPI H-band spectroscopy reported in \citet{greenbaumpueyo2018}. We do not take the spectral covariance into account for GPI or SPHERE. For both SPHERE and GPI we  fit for a scaling factor with respect to the GRAVITY observation during the retrieval, as was also done in \citet{nowaklacour2020}. This also appears necessary given the noticeable shift between the SPHERE and GPI observations in their overlapping region at $\sim 1.6$ micron. Similar shifts between GPI and SPHERE observations have been reported in \citet{samlandmolliere2017}. However, this difference could also be caused by variability, because spectral template brown dwarfs that reproduce the spectral properties of HR~8799e well have been found to exhibit considerable variability, possibly up to a 20-30\% peak-to-peak amplitude (see the discussion of \citealt{macekirkpatrick2013,billervos2015} in \citealt{bonnefoyzurlo2016}).

\subsubsection*{Archival photometry}
Although not included during the fit, we compare our results with archival photometry in the mid-infrared. We consider the 3.3~$\mu$m LBT photometry reported in \citet{skemerhinz2012}, the L' band and [4.05]-Br$\alpha$ photometry reported in \citet{currieburrows2014}, as well as the M' band upper limit of \citet{galichermarois2011}. The photometry was converted from magnitudes to flux using the \emph{species}\footnote{\url{https://species.readthedocs.io}} toolkit, which has been described in \citet{stolkerquanz2019}. \rch{We decided against including the photometric fluxes in the fit because their relatively low signal-to-noise would add little constraining power to the retrieval, when compared to the spectra, but would double the run-time of our retrievals due to the increased spectral range.}

\begin{table}[t!]
\centering
{ \footnotesize
\begin{tabular}{ll|ll}
\hline \hline
Parameter & Prior & Parameter & Prior \\ \hline
$T_1$ & $\mathcal{U}(0 , T_2)$ & ${\rm log}(\tilde{X}_{\rm Fe})^{\rm (c)}$ & $\mathcal{U}(-2.3, 1)$ \\
$T_2$ & $\mathcal{U}(0 , T_3)$ & ${\rm log}(\tilde{X}_{\rm MgSiO_3})$ & $\mathcal{U}(-2.3, 1)$ \\
$T_3$ & $\mathcal{U}(0 , T_{\rm connect})^{\rm (a)}$ & $f_{\rm sed}$ & $\mathcal{U}(0, 10)$ \\
${\rm log}(\delta)$ & $P_{\rm phot} \in [10^{-3}, 100]^{\rm (b)}$ & ${\rm log}(K_{\rm zz})$ & $\mathcal{U}(5, 13)$ \\
$\alpha$ & $\mathcal{U}(1,2)$ & $\sigma_{\rm g}$ & $\mathcal{U}(1.05, 3)$ \\
$T_0$ & $\mathcal{U}(300 , 2300 )$ & $R_{\rm P}$ & $\mathcal{U}(0.9, 2)$ \\
$\rm C/O$ & $\mathcal{U}(0.1,1.6)$ & ${\rm log}(g)$ & $\mathcal{U}(2, 5.5)$ \\
$\rm [Fe/H]$ & $\mathcal{U}(-1.5,1.5)$ & ${\rm log}(P_{\rm quench})$ & $\mathcal{U}(-6, 3)$ \\ \hline
$f_{\rm SPHERE}$ & $\mathcal{U}(0.8,1.2)$ & $f_{\rm GPI}$ & $\mathcal{U}(0.8,1.2)$ \\ \hline
\end{tabular}
}
\caption{Priors of the HR~8799e retrieval. $\mathcal{U}$ stands for a uniform distribution, with the two parameters being the range boundaries. The units for the parameters are the same as the ones used for Table \ref{tab:ret_input}. $f_{\rm SPHERE}$ and $f_{\rm GPI}$ are the scaling factors retrieved for the SPHERE and GPI data, respectively. (a) and (b): please see Section \ref{sect:ret_model_setup} for a definition of $P_{\rm phot}$ and $T_{\rm connect}$. It holds that $\tilde{X}_i = X^{\rm i}_0/X^{\rm i}_{\rm eq}$, where the latter quantity has been defined in Table \ref{tab:ret_input}.}
\label{tab:samp_prior}
\end{table}

\subsection{Retrieval model setup}
\label{sect:ret_model_setup}
We set up our nominal retrieval model with 18 free parameters, using Cloud Model 1, which we describe in the following. Additionally, we will compare to retrievals using Cloud Model 2, see Section \ref{sect:hr8799_cm2}. The free parameters and prior ranges are listed in Table \ref{tab:samp_prior}. As in \citet{nowaklacour2020}, we fitted multiplicative scaling factors $f_{\rm SPHERE}$ and $f_{\rm GPI}$ to account for systematic biases in the flux normalization of these datasets. The $T_{\rm connect}$ quantity referenced in the prior range of $T_3$ is the uppermost temperature of the `photospheric' layer, and was calculated by setting $\tau=0.1$ in Equation \ref{equ:t_edd}. Like the priors for $T_2$ and $T_1$, this ensures a temperature profile that is monotonically decreasing with altitude, also see Section \ref{sect:priors}. $\delta$ was sampled from the prior by assuming a log-uniform prior on $P_{\rm phot}$, where we defined $P_{\rm phot}$ as the pressure where $\tau=1$ in Equation \ref{equ:tau_pt}. This allowed to solve for $\delta$ for a given $P_{\rm phot}$ value. The following absorber species were included: CO, CO$_2$ and H$_2$O \citep[from][]{rothman2010}, CH$_4$ \citep{yurchenko2014}, NH$_3$ \citep{yurchenkobarber2011}, H$_2$S \citep{rothman2013}, Na and K (\citealt{piskunov1995}, with Allard wings, see \citealt{mollierewardenier2019} for more details), PH$_3$ \citep{sousa-silvaal-refaie14}, VO and TiO (Plez line lists, see \citealt{mollierewardenier2019} for more details), FeH \citep{wendereiners2010} as line absorbers, H$_2$, He as Rayleigh scatterers, the collision induced absorption of H$_2$-H$_2$, H$_2$-He, and the scattering and absorption cross sections of crystalline, irregularly shaped Fe and \ce{MgSiO3(c)} cloud particles. See \citet{mollierewardenier2019} for the full reference list and description of the opacity sources. The FeH opacity has been multiplied by a factor $1/2$ due to the partition function correction described in \citet{charnaybezard2018}. We convolved the synthetic spectra using a Gaussian kernel, in order to approximate the line spread function of the SPHERE, GPI, and GRAVITY instruments. The instrumental resolving power was assumed to be 30, 45 and 500 for SPHERE, GPI and GRAVITY, respectively. The resolution element $\Delta \lambda$ of the spectrograph was assumed to be the FWHM of the line spread function. This means that the standard deviation $\sigma_{\rm LSF}$ of the Gaussian kernel used for convolution is defined by $\Delta \lambda = 2\sqrt{2 \ {\rm ln}2}\ \sigma_{\rm LSF}$. The retrieval was run with PyMultiNest, using 4000 live points.

\subsubsection*{Pressure grid}
We use an adaptive spacing for the atmosphere's pressure grid. For cloud-free calculations the atmosphere would be separated into 60 points, spaced equidistantly in log-pressure between $10^{-6}$ and 1000~bar. For cloudy calculations the retrieval code considers all cloud base pressures $P_{\rm base}$ and increases the spatial resolution for $P\in [0.5 \ P_{\rm base}, 1.12 \ P_{\rm base}]$ (corresponding to a pressure range of -0.3 and 0.05 dex) by a factor of 12. This better resolves the placement of the cloud base within the atmosphere, in case a $P_{\rm base}$ does not fall into the immediate proximity of a grid point of the coarse pressure grid. In addition, the abrupt increase in atmospheric opacity at the cloud deck position, and its decline $\propto P^{f_{\rm sed}}$ for lower pressures, is resolved better. For two spatially separated cloud decks this leads to 104 grid points. We found that this treatment is as accurate as running the whole calculation at a 12 times-increased resolution, which would result in 720 grid points. For this we compared to a baseline calculation made at a 24 times higher resolution, using 1440 grid points. 

\subsubsection*{Testing the retrieval model}
\label{sect:test_hr8799e_model}
Because all test retrievals mentioned in Section \ref{sect:ver_ret} were carried out on data sets of homogeneous wavelength coverage in YJHK bands, we verified our retrieval setup by running a mock retrieval that had the same wavelength spacing and error properties as the actual HR8799e data sets of SPHERE, GPI, and GRAVITY. As input for the synthetic observation we used a posterior sample of the actual HR8799e retrieval, the result is shown in Figure \ref{fig:combine_all_ver_ret_like_real_data}. 
We find that we can retrieve all parameters well, except for [Fe/H] \rch{and $K_{zz}$}, which are biased, as are the scaling parameters. Running a second fit that neglected the scaling parameters lead to a well retrieved [Fe/H], but slightly too small radius. This behavior could be due to the random noise instantiation used in the retrieval, and the fact that especially the scaling may introduce biases in the retrieved atmospheric parameters in cases where differences are introduced between model and observation. This has also been described in \citet{kitzmannheng2019}. In their case the differences arose from using two different models for generating the mock observation and the retrieval, here the differences arise from the random noise properties. We note that the scaling value we retrieve for the actual HR8799e data below is consistent with unity.

\rch{To test the impact of the random noise instantiation further we also ran test retrievals for the same synthetic observation, but neglecting the random perturbation of the data due to the noise. Similar to our noise-free test retrieval presented in Section \ref{sect:cloud_model1_test_retrieval}, we found that the noise-free test led to a bi-modal posterior, with the modes bracketing the input values. An analogous approach (zooming in on the prior volume populated by the bi-modal posterior) lead to a uni-modal posterior, retrieving the input parameters. Thus we reconfirm our observation that noiseless test retrievals can lead to multi-modal posteriors if the ratio of the prior volume and the number of live points is large. This indicates that multiple parameter combinations can lead to excellent spectral fits, while within 1~$\sigma$ of the true parameters. As stated in the manuscript before this may indicate that a retrieval model with fewer free parameters could be favored, leading to more unique solutions. We note that in these test retrievals described here, the atmospheric C/O ratio was a robustly retrieved parameter in all retrieval setups, and that our two retrievals for the real HR~8799e data presented below, using either Cloud Model 1 or 2, lead to consistent C/O, [Fe/H] and ${\rm log}(g)$ constraints.}

\begin{figure*}[t!]
\centering
\includegraphics[width=0.88\textwidth]{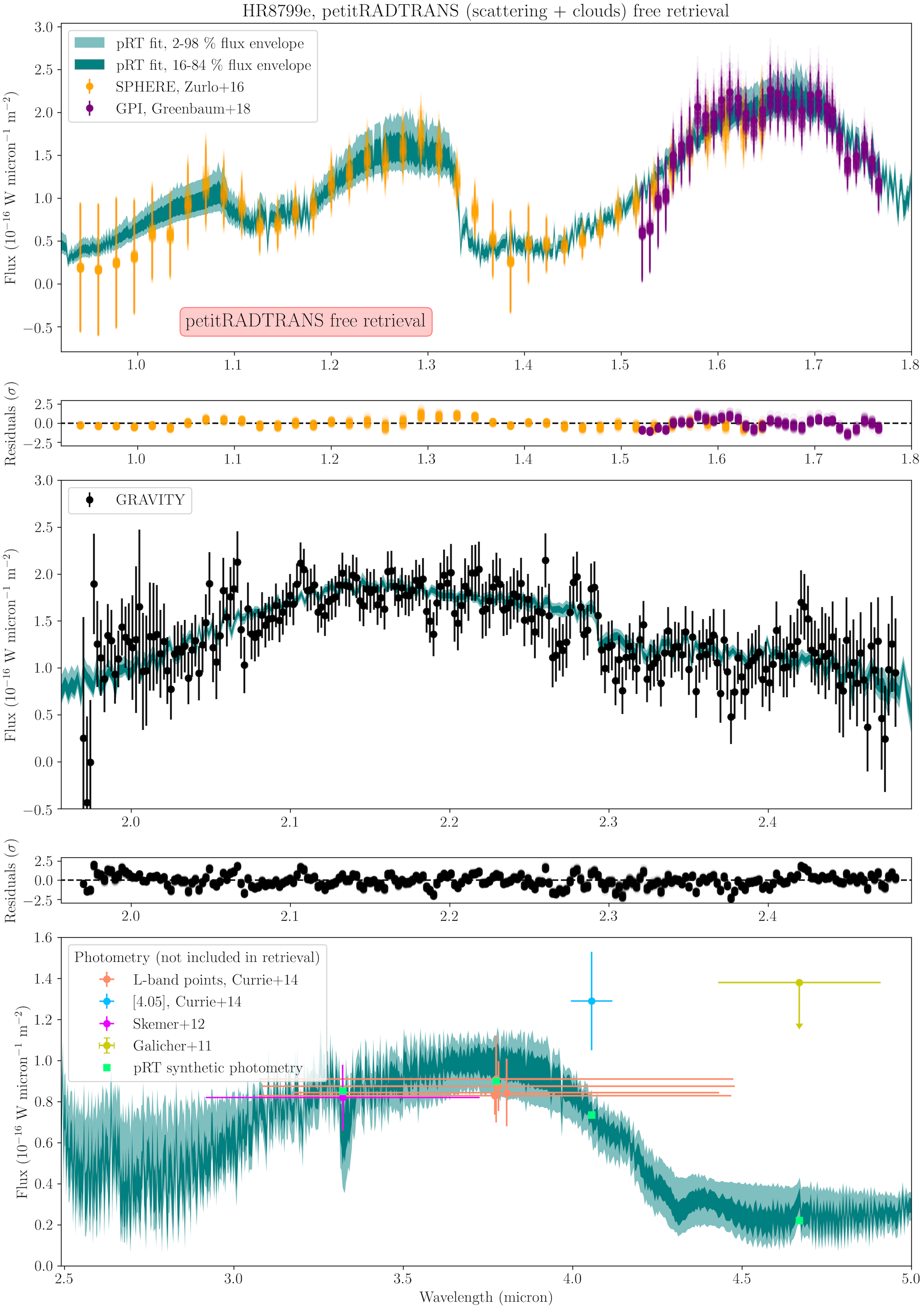}
\caption{Spectral fit of HR~8799e. The {\it upper panel} shows the YJH-band observations of SPHERE and GPI, the {\it middle panel} the GRAVITY K-band observations. The {\it lowest panel} shows the photometry of the planet, which was not included during the retrieval. The 16-84 and 2-98 \% flux envelopes of the sampled \ptrad \ retrieval models are shown in all panels. Because also the SPHERE and GPI scaling factors were sampled 100 times for making this plot, there are multiple points visible at every wavelength.}
\label{fig:HR8799retrieval_spec}
\end{figure*}

\begin{figure*}[t!]
\centering
\includegraphics[width=0.48\textwidth]{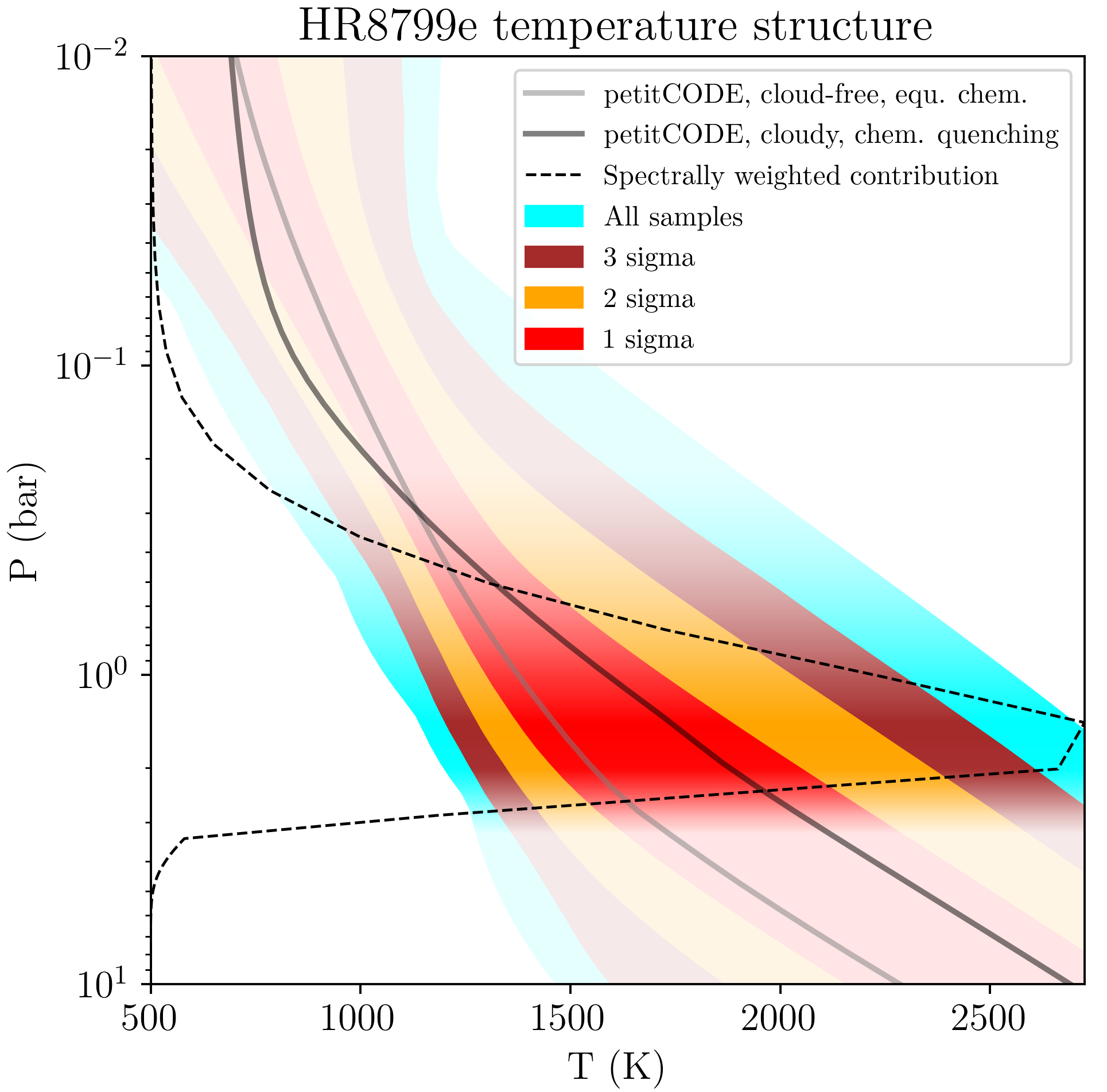}
\includegraphics[width=0.48\textwidth]{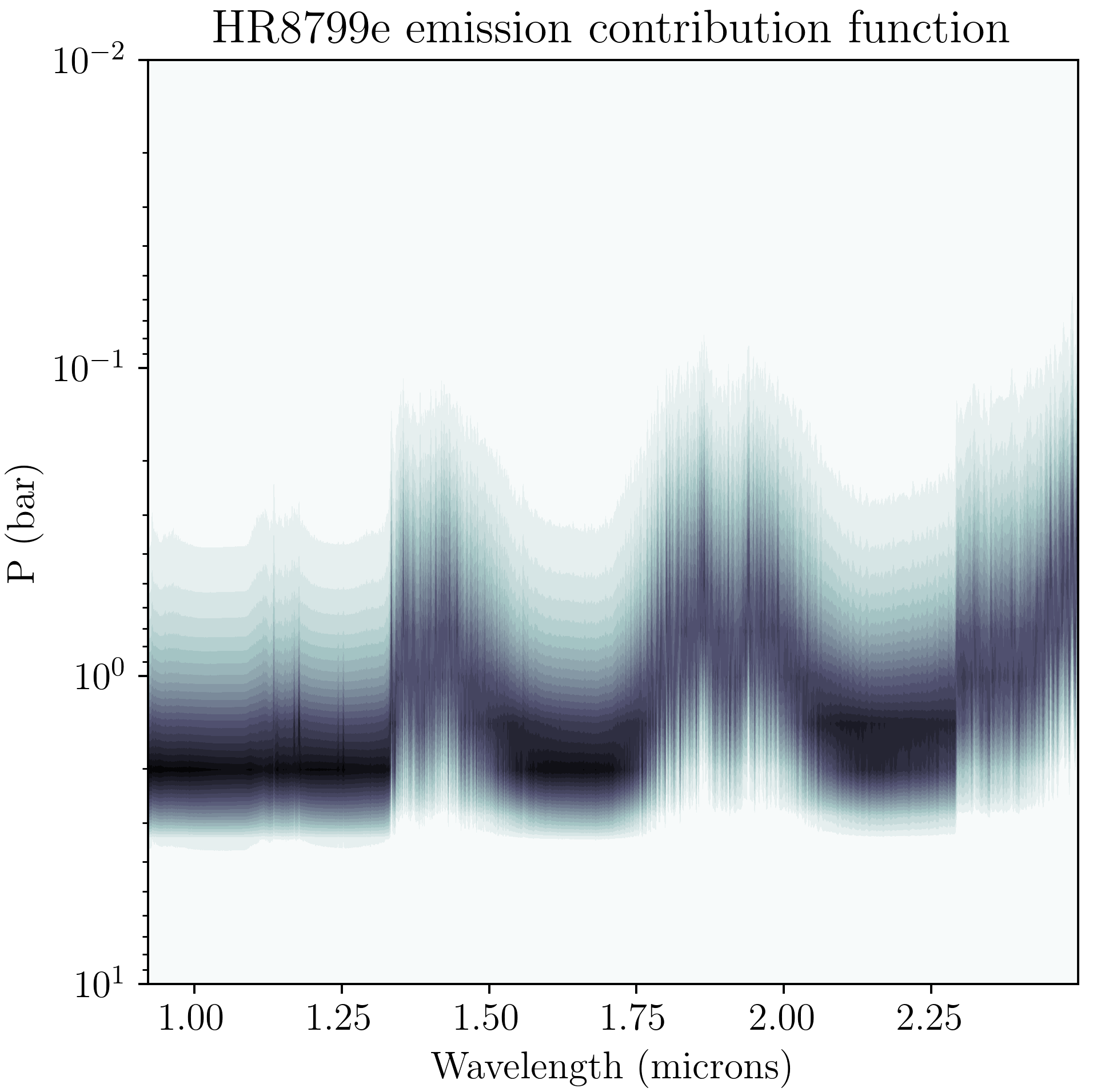}
\caption{{\it Left panel}: temperature distribution of the atmosphere of HR8799e, retrieved with the \ptrad \ free retrieval setup. See the caption of Figure \ref{fig:combine_all_alpha_12_1000_noise_free} for an explanation of how to read this plot. In addition the self-consistent $P$-$T$ curves derived from \emph{petitCODE}, assuming chemical equilibrium and no clouds, or chemical quenching with clouds, are shown as gray and black solid lines, respectively. {\it Right panel}: emission contribution function of the best-fit model of the HR8799e retrieval.}
\label{fig:HR8799retrieval_PTs_contribution}
\end{figure*}

\subsection{Free retrieval results of HR8799e}
\label{sect:ret_hr8799_for_real}
In this section we describe the retrieval results. The results will also be discussed in view of the possible planet formation history, and compared to existing literature studies of HR8799e in sections \ref{sect:formation} and \ref{sect:literature_comp}, respectively.

The spectral fit for HR8799e is presented in Figure \ref{fig:HR8799retrieval_spec}. In general, the retrieval model is able to explain the observations well. The residuals scatter around zero, with some systematic differences visible at 1.325, 1.525, 1.725, 2.07, 2.11, 2.18, 2.275 and 2.425~micron. These differences could be due to the model missing absorbers, or not being flexible enough to fit intricacies in the atmospheric temperature, abundance, or cloud structure. Another likely possibility are remaining systematics in the observations: the difference between the SPHERE and GPI observations in their overlap region (1.525~micron), as well as the overall wiggly appearance of the GRAVITY observation may indicate this. We note that also the photometric flux measurements in the MIR are fit well, except for the narrow [4.05] band measurement by \citet{currieburrows2014}.\footnote{We used \emph{species} to convert the \ptrad \ flux to photometric fluxes.} This is especially interesting as these points were not included in the retrieval. Especially the 3.3~$\mu$m LBT - L' and L'-[4.05] colors have been noted to be difficult to explain with self-consistent models, see discussion in Section \ref{sect:literature_comp}.

From the spectral appearance of the H and K band observations it is already clear that \ce{CH4} is not an important absorber in the atmosphere: the flux decrease expected from \ce{CH4} absorption at 1.6 and 2.2~micron is absent. The fact that only an upper limit is found by observations in the M-band, together with the comparatively high flux in the 3.3~$\mu$m LBT and L' bands, also speaks for a \ce{CH4}-poor atmosphere.

\begin{figure*}[t!]
\centering
\includegraphics[width=0.95\textwidth]{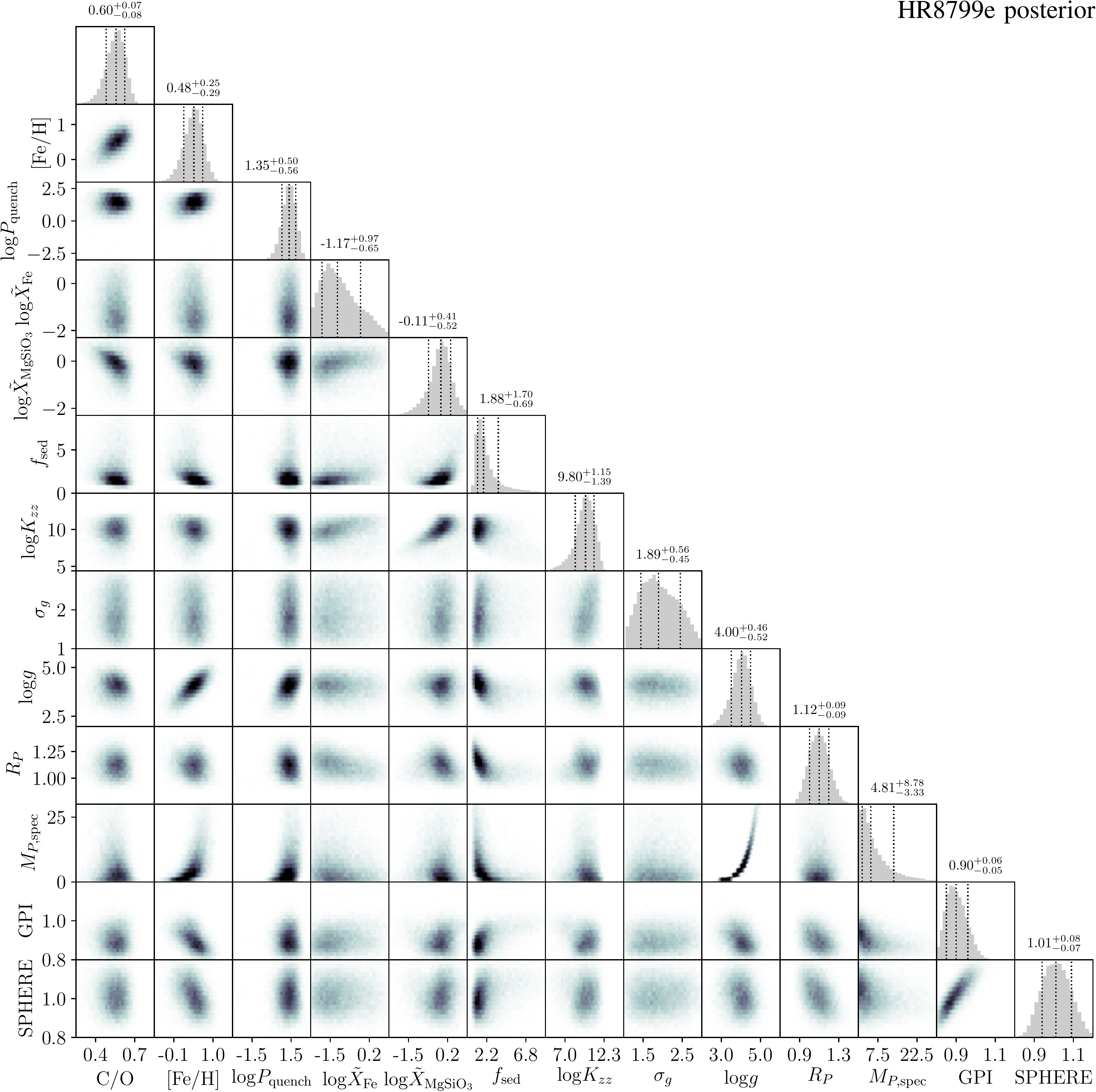}
\caption{Corner plot of the free retrieval of HR~8799e with \ptrad. This plot shows the one and two-dimensional projections of the 18-dimensional posterior distribution for all but the six nuisance parameters of the temperature structure (see the left panel of Figure \ref{fig:HR8799retrieval_PTs_contribution} for the P-T confidence envelopes). $M_{\rm P,spec}$ is the mass of HR~8799e in units of Jupiter masses, derived from the retrieval posterior distributions of ${\rm log}(g)$ and $R_{\rm P}$.}
\label{fig:HR8799retrieval_corner}
\end{figure*}

The retrieved pressure-temperature uncertainty envelopes of HR8799e are shown in the left panel of Figure \ref{fig:HR8799retrieval_PTs_contribution}. As in the analogous plot shown in Figure \ref{fig:combine_all_alpha_12_1000_noise_free}, the opacity of the uncertainty envelopes is scaled by the flux-averaged emission contribution function of the best-fit model, to show where in the atmosphere the observations are probing. This contribution function is also shown in the right panel of Figure \ref{fig:HR8799retrieval_PTs_contribution}. For comparison, we also show the P-T curves derived with our self-consistent \petit using the best-fit parameters of the retrieval. This is further discussed in Section \ref{sect:ptcode_pts_comp}.

The corner plot of the one- and two-dimensional projections of the 18-dimensional posterior distribution of the retrieval is shown in Figure \ref{fig:HR8799retrieval_corner}. We summarize a few of the most striking results here, while the implications of the retrieved parameter values will be discussed in sections \ref{sect:formation} and \ref{sect:literature_comp}. In general, we note that none of the retrieved parameters ran into its prior boundaries and that all parameters (except for the Fe mass fraction at the cloud base) are well constrained. In addition, we derive that the atmospheric C/O ratio is ${\rm C/O} = 0.60_{-0.08}^{+0.07}$ and the retrieved metallicity is ${\rm [Fe/H]} = 0.48_{-0.29}^{+0.25}$. Together with the planet's mass, which we derive to be $4.81_{-3.33}^{+8.78} \ \mj$, this has important implications for how the planet could have formed, see Section \ref{sect:formation}. We note that the surface gravity and radius retrieved for HR8799e, ${\rm log}(g)=4.00_{-0.52}^{+0.46}$ and $R_{\rm P}=1.12_{-0.09}^{+0.09}\ \rj$, are constrained with a symmetric, uni-modal peak. Hence also the logarithm of the planet mass is constrained with a symmetric, uni-modal peak, but the distribution of the mass itself is skewed towards lower masses, with large mass uncertainties due to the large uncertainty on ${\rm log}(g)$. The atmosphere is clearly affected by disequilibrium chemistry, with a large quench pressure of ${\rm log}(P_{\rm quench}/{\rm 1 \ bar})=1.35_{-0.56}^{+0.50}$. We find that disabling quenching at the best-fit parameters leads to strong \ce{CH4} absorption features in the spectrum, which are inconsistent with the data. Moreover, the scaling value retrieved for SPHERE is consistent with one, $1.01_{-0.07}^{+0.08}$. The scaling retrieved for GPI is significantly smaller than 1, namely $0.90_{-0.05}^{+0.06}$. Only when applying this scaling on GPI do the SPHERE and GPI datasets agree in their overlapping region, see Figure \ref{fig:HR8799retrieval_spec}. From sampling the posterior distribution 300 times, and calculating the spectra between 0.5 and 28~micron we derive an effective temperature of $T_{\rm eff}=1154_{-48}^{+49}$~K.
Using this derived temperature, our retrieved surface gravity, and Equation (4) of \citet{zahnlemarley2014}, we find that the upper limit for the atmospheric mixing is ${\rm log}(K_{zz,{\rm max}})=10.2$, which assumes that all flux is transported by convection. Our derived value, which is used solely for determining the particle size for a given $f_{\rm sed}$, is ${\rm log}(K_{zz})=9.80_{-1.39}^{+1.15}$, so below the theoretical upper limit (while still large).

\subsection{Retrieval with the non-nominal Cloud Model 2}
\label{sect:hr8799_cm2}

To test the robustness of the constrained atmospheric properties, we also retrieved HR8799e with Cloud Model 2. Like before, we find that Cloud Model 2 leads to retrieved temperature gradients which are too shallow when compared to physically consistent solutions. The retrieved atmospheric solution is bi-modal, with one solution corresponding to a P-T structure with a shallow temperature gradient and intermediate cloudiness, and a second solution corresponding to an even more isothermal, cloud-free atmosphere. The spectral fit, full posterior distribution, and P-T uncertainty envelopes are shown in Appendix \ref{sect:app_hr8799e_CM2}, where we also describe the prior setup of this retrieval.

\begin{figure}[t!]
\centering
\includegraphics[width=0.48\textwidth]{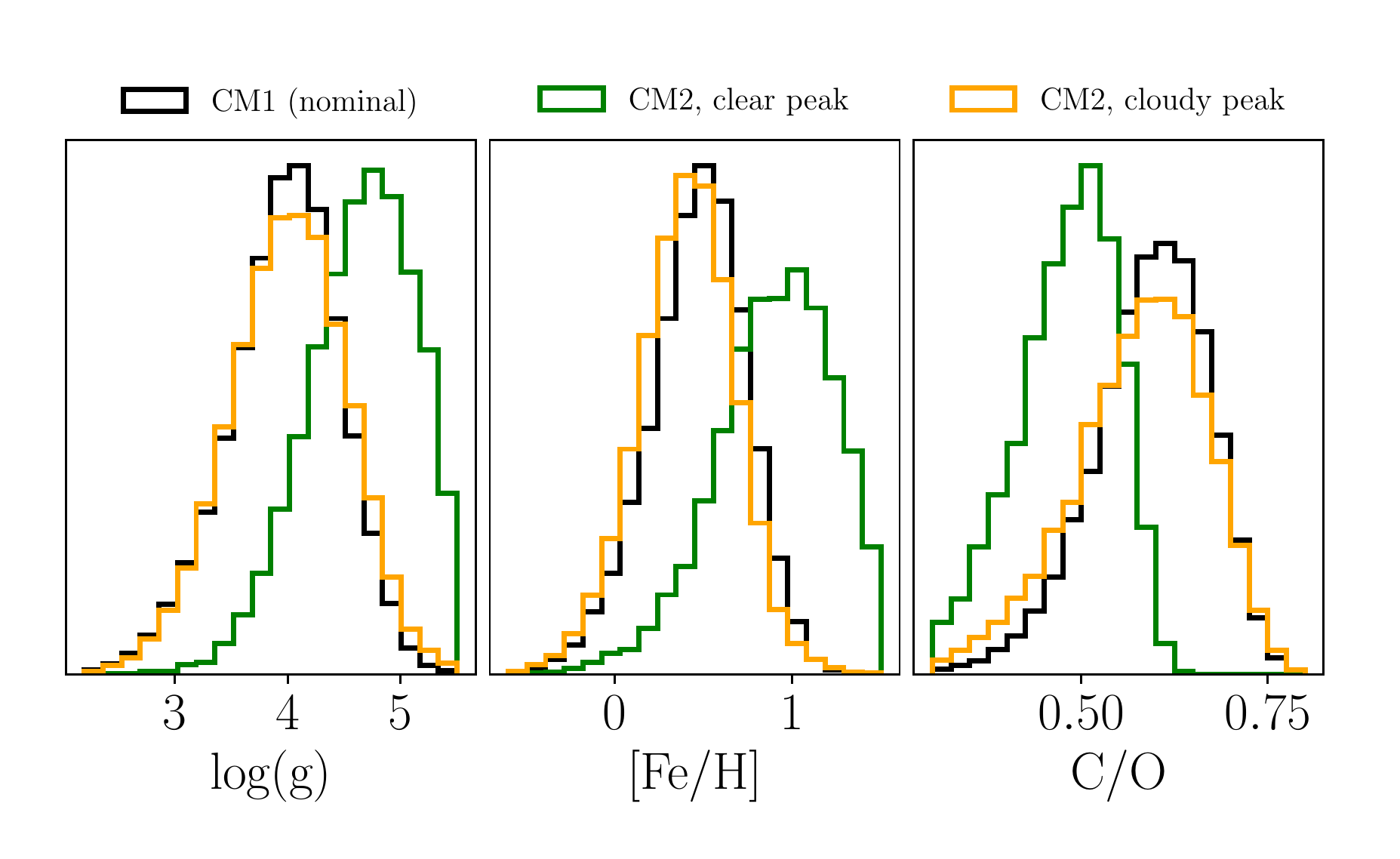}
\caption{Marginalized one-dimensional posterior distributions of HR8799e's gravity, metallicity, and C/O, shown for our nominal retrieval with Cloud Model 1 (black solid line), and the retrieval with Cloud Model 2, which lead to a bi-modal solution of a cloudy (orange solid line) and $\sim$isothermal, clear (green solid line) atmospheric state.}
\label{fig:HR8799CM1CM2}
\end{figure}

Focusing on the bulk atmospheric properties (${\rm log}(g)$, C/O, [Fe/H]), we find that the cloudy mode of the retrieval with Cloud Model 2 is fully consistent with the results from our nominal retrieval with Cloud Model 1. The one-dimensional posterior distributions look almost identical. Thus, even though the atmospheric temperature and cloud structure retrieved with different cloud models can differ, {\it quantities such as C/O, [Fe/H] and the atmospheric gravity may be very robust}. The clear, isothermal solution of Cloud Model 2 leads to different values for these retrieved parameters, but their 1-$\sigma$ uncertainty regions overlap. Figure \ref{fig:HR8799CM1CM2} shows the marginalized one-dimensional posterior distributions of HR8799e's gravity, metallicity, and C/O, derived with Cloud Model 1 and Cloud Model 2.

\subsection{Comparison of the results with self-consistent atmospheric models}
Free retrievals allow to deviate from the rigidity of (potentially imperfect) physical assumptions in self-consistent codes. This can be both boon and bane. On one hand it allows accounting for effects that influence the atmospheric structure which are not adequately captured by the physical assumptions made in the self-consistent codes. On the other hand, the free retrieval may converge on parameter results which lead to a seemingly good fit to the data, but are unphysical. Because of this, it is crucial to verify retrieval results by comparing to constraints that can be obtained from self-consistent codes, as was also done in, for example, \citet{linemarley2017,gandhimadhusudhan2018}. We describe such tests below: we compared our derived atmospheric structure with self-consistent results and compared our free retrieval to a grid interpolation retrieval with self-consistent atmospheric spectra.

\subsubsection*{Self-consistent P-T structures of \petit}
\label{sect:ptcode_pts_comp}
\petit \citep{mollierevanboekel2015,mollierevanboekel2016} is a self-consistent code for calculating atmospheric structures and spectra. It assumes radiative-convective and chemical equilibrium to calculate the atmospheric structure. The radiative transfer includes scattering, and the code can include gas line, continuum, and cloud opacities.

We carried out two tests: for the overall planet parameters (${\rm log}(g)$, $T_{\rm eff}$, [Fe/H], C/O), we used the median of the retrieved values. We then calculated a cloud-free atmospheric structure, assuming chemical equilibrium. As a second test we included clouds, prescribed the median values of the retrieved cloud parameters ($f_{\rm sed}$, $K_{zz}$, $\sigma_g$, $\tilde{X}_{\rm Fe}$, $\tilde{X}_{\rm MgSiO3}$), and enforced that the \ce{H2O}, \ce{CH4} and CO abundances be held constant below the retrieved best-fit quench pressure. The resulting structures are shown as gray and black solid lines in the left panel of Figure \ref{fig:HR8799retrieval_PTs_contribution} for the cloud-free chemical equilibrium and the cloudy non-equilibrium structure, respectively.

Overall, it can be seen that the self-consistent structures follow the uncertainty envelopes of the retrieved pressure temperature structure well, in terms of absolute temperature and slope. Interestingly, we notice that the cloud-free structure in chemical equilibrium falls within the 1-$\sigma$ envelope at all pressures, while the self-consistent cloudy structure which included quenching moves out into the 2-$\sigma$ envelope between 0.05 and 0.3 bar. This is above the region of maximum emission as measured by the contribution function of the best-fit model, however. We conclude that we do not see any clear deviation of our retrieved temperature envelopes when compared to physical expectations, both the overall shape and absolute temperatures appear to be close to what is predicted in a self-consistent model.

\subsubsection*{Spectral fit with Exo-REM}
\label{sect:exo_rem_fit}

In addition to the free retrieval with \ptrad \ described above, we carried out a grid-interpolation retrieval to obtain the atmospheric parameters of HR8799e from its spectrum. The grid of self-consistent model spectra was obtained with \emph{Exo-REM} \citep{baudinobezard2015,baudinomolliere2017,charnaybezard2018}, in the version by \citet{charnaybezard2018}, which includes scattering and disequilibrium chemistry. The chemical disequilibrium and cloud scale height is determined through taking into account the vertical atmospheric mixing, which is set through the atmospheric eddy diffusion parameter $K_{\rm zz}$. In \emph{Exo-REM}, $K_{\rm zz}$ is determined from the atmospheric structure in the  convective region consistently, using mixing length theory. Above the convective region, $K_{\rm zz}$ is determined from a convective overshooting description. \emph{Exo-REM} includes the cloud opacities of spherical, amorphous Fe and \ce{Mg2SiO4} grains, as well as the gas opacities of Na, K, H$_2$O, CH$_4$, CO, CO$_2$, NH$_3$, PH$_3$, TiO, VO, and FeH, in addition to H$_2$-H$_2$ and H$_2$-He collision-induced absorption (CIA).
The \emph{Exo-REM} grid used here ranged in $T_{\rm eff}$ from 1000-2000~K ($\Delta T_{\rm eff}=50$~K), and in C/O from 0.3-0.75 ($\Delta{\rm C/O}=0.05$). The [Fe/H] grid points were $-$0.5, 0, 0.5, while the ${\rm log}(g)$ points were at 3.5, 4, 4.5.

Notable points of difference in the opacity treatment between \emph{Exo-REM} and \ptrad \ are the use of different alkali wing profiles (\citealt{burrows2003} for \emph{Exo-REM}, \citealt{allard2003,allardkielkopf2012} for \ptrad) and that \ptrad \ assumed irragularly shaped, crystalline Fe and \ce{MgSiO3} cloud particles instead of spherical, amorphous Fe and \ce{Mg2SiO4} ones. Moreover, \emph{Exo-REM} varies C/O by changing C, whereas \ptrad \ varies C/O by changing O, which can make a difference \citep{lodders2010}.

\begin{figure*}[t!]
\centering
\includegraphics[width=0.88\textwidth]{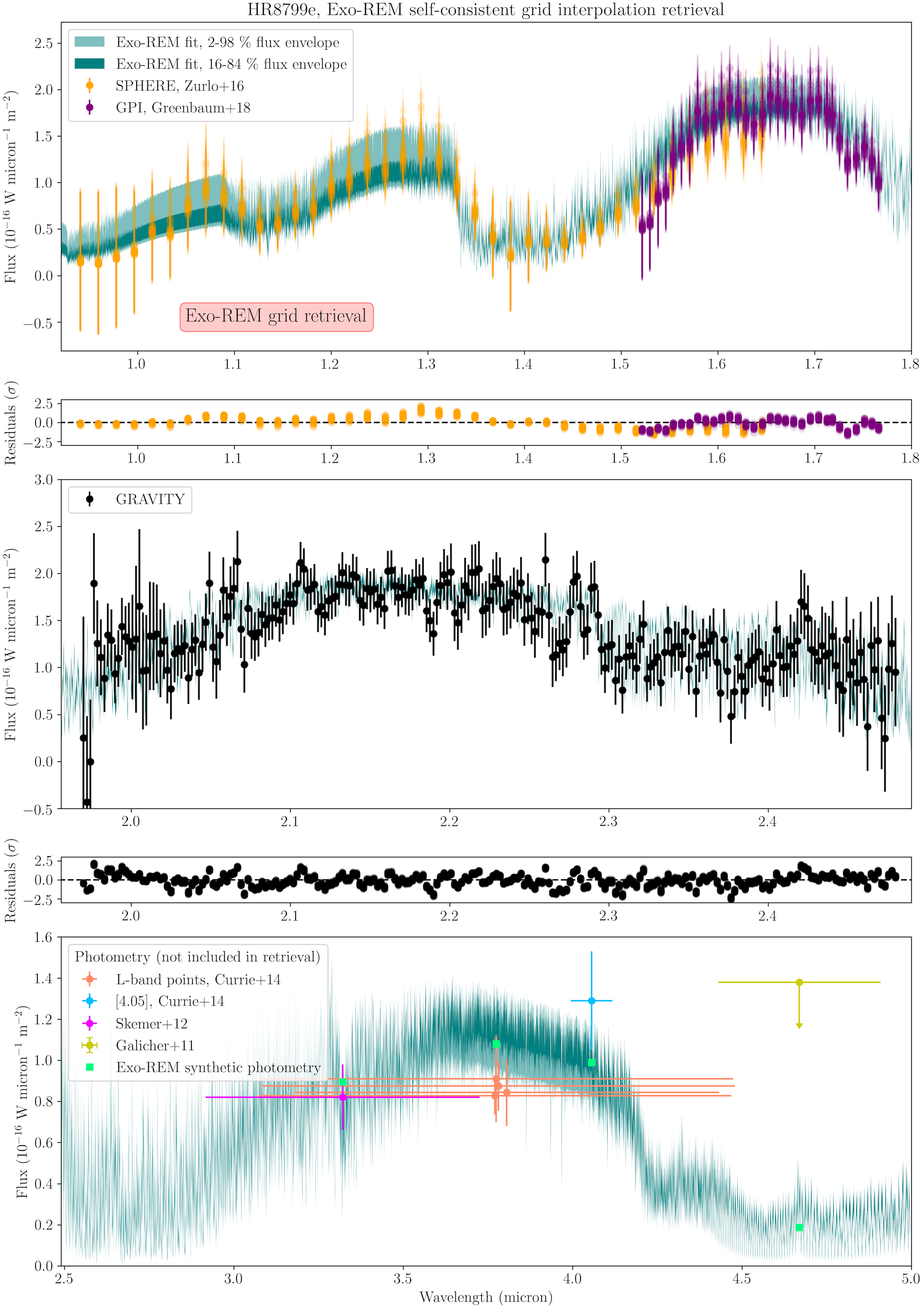}
\caption{Same as Figure \ref{fig:HR8799retrieval_spec}, but for a grid-based interpolation retrieval with the self-consistent code \emph{Exo-REM}.}
\label{fig:HR8799retrieval_spec_Exo-REM}
\end{figure*}

Applying a LSF convolution and rebinning identically to the \ptrad \ retrieval, we used \emph{species} to carry out a grid-based retrieval with \emph{PyMultiNest}, where the model spectra were obtained by linearly interpolating within the \emph{Exo-REM} grid. 100 spectra sampled from the Exo-REM posterior are shown in Figure \ref{fig:HR8799retrieval_spec_Exo-REM}, analogous to the spectra shown for the free retrieval with \ptrad \ in Figure \ref{fig:HR8799retrieval_spec}.

\begin{figure*}[t!]
\centering
\includegraphics[width=0.99\textwidth]{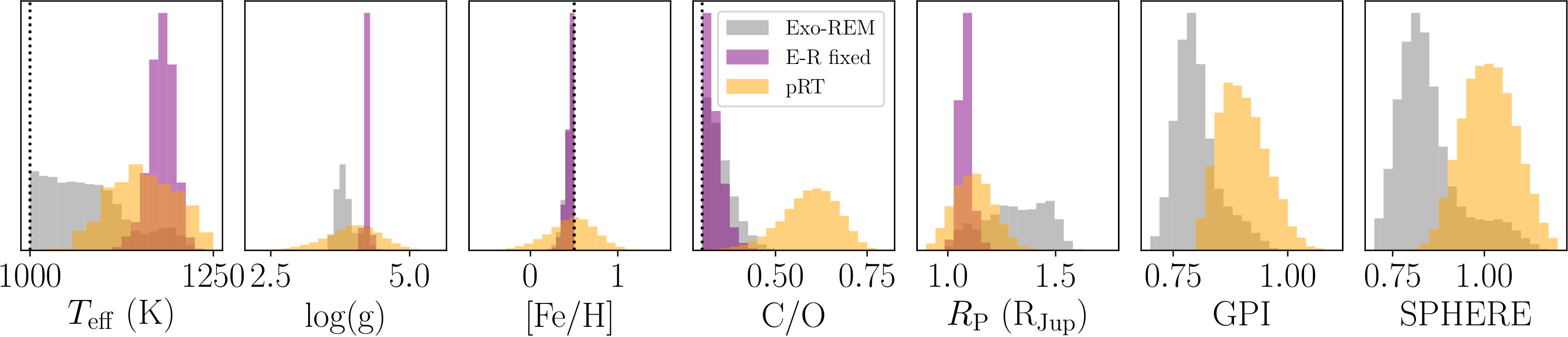}
\caption{\rch{One-dimensional marginalization of the HR~8799e retrieval posteriors for the self-consistent \emph{Exo-REM} grid when retrieving the SPHERE/GPI flux scaling (gray), or fixing it to the median values of \ptrad \ (purple). The results of the freely parameterized \ptrad \ retrieval are shown in orange.} The vertical dashed lines denote the upper [Fe/H], and lower C/O and $T_{\rm eff}$ grid boundaries of the \emph{Exo-REM} models.}
\label{fig:Exo-REM_pRT_posterior_comp}
\end{figure*}

In general the quality of the spectral fit is very similar to the free retrieval with \ptrad, with the difference that the residuals in the near-IR are a somewhat larger. Like \ptrad, \emph{Exo-REM} can fit the photometry in the mid-IR well, even though it was not included in the fit. Analogous to the \ptrad \ fit, also \emph{Exo-REM} does not reproduce the [4.05] band photometric point well, but is somewhat more consistent than \ptrad. 

We show the one-dimensional marginalizations of the \emph{Exo-REM} posterior distribution in Figure \ref{fig:Exo-REM_pRT_posterior_comp}, together with the marginalized posteriors of the corresponding parameters of the free \ptrad \ retrieval. Because the effective temperature $T_{\rm eff}$ is not a free parameter of the \ptrad \ retrieval, the $T_{\rm eff}$ distribution was obtained from sampling the posterior 300 times, and calculating the effective temperature from the spectrum ranging from 0.5 to 28~$\mu$m.  \rch{We find two modes in the \emph{Exo-REM} retrieval. The first mode has lower temperatures, smaller GPI and SPHERE scaling factors, and larger planet radii when compared to the \ptrad \ retrieval. For the second mode the scaling factors, $T_{\rm eff}$ and radius are consistent with the \ptrad \ results. The surface gravity agrees with the broad \ptrad \ posterior, for both modes. The lower temperature mode of the \emph{Exo-REM} fits runs into the lower grid boundary for the effective temperature. Both modes run into to lower grid boundary for C/O, and in the upper grid boundary for [Fe/H]. Due to the inherent difficulty of converging self-consistent cloudy models the \emph{Exo-REM} grid is incomplete, that is, models in the grid are missing especially at lower temperatures. This could affect the fit at low effective temperatures. Therefore, as a second test, we fixed the GPI and SPHERE scaling factors to the best-fit values of the \ptrad \ fit, which is also shown in Figure \ref{fig:Exo-REM_pRT_posterior_comp}. This leads to effective temperatures and radii consistent with the \ptrad \ fit, while still running into the C/O and [Fe/H] boundaries.}

\rch{For both the metallicity and C/O ratio the comparison between the \ptrad \ and \emph{Exo-REM} results is difficult: the retrieved best-fit metallicity value from the \emph{Exo-REM} grid is consistent with the \ptrad \ peak metallicity. However, the \emph{Exo-REM} retrieval runs into the upper boundary of the grid. For the C/O ratio the \ptrad \ fit peaks at around 0.6, while it is driven into the lower grid boundary (0.3) for the \emph{Exo-REM} fit. The behavior of the \emph{Exo-REM} posterior for these two quantities therefore makes a more detailed comparison of the \ptrad \ and \emph{Exo-REM} retrievals difficult. It appears as if the \emph{Exo-REM} grid may not be able to fully reproduce the near-IR photometry, and trends into the boundaries of the grid in search of the true probability maximum that is not contained within its grid boundaries.}

\rch{We can speculate at the near-IR region being the cause for the difficulties we face in the \emph{Exo-REM} retrieval: here the residuals are larger than in the \ptrad \ retrieval. Because the YJH bands are the most strongly affected by clouds, this could hint at a difference in the description of clouds, but also the alkalis could play a role, especially in the SPHERE Y band. Indeed, when only retrieving the atmospheric properties using the GRAVITY K-band with \emph{Exo-REM}, we find that the C/O is constrained at $0.43_{-0.07}^{+0.07}$, while [Fe/H] still trends into the upper grid boundary (0.5). A larger grid extend with additional [Fe/H] and C/O values, may therefore alleviate the problems we face in our analysis here. However, we expect that a grid that also varies the cloud parameters of \emph{Exo-REM} may help, as we find that decreasing the C/O (or increasing the metallicity) leads to stronger water absorption features across the YJH bands. This speaks for C/O being used to counteract too strong cloud absorption.}

We conclude that the \ptrad \ and \emph{Exo-REM} retrievals are consistent with each other in the parameters that can be easily compared. However, parameters trending into the grid boundary in the \emph{Exo-REM} retrieval make the comparison difficult for [Fe/H] and C/O.

\section{Discussion}
\label{sect:discussion}
\subsection{Implication of the retrieved C/O and [Fe/H] for the formation location of HR~8799e}
\label{sect:formation}

Measuring a planet's C/O has been suggested as a powerful tool to trace where in the protoplanetary disk a planet may have formed \citep[e.g.,][]{oeberg2011,madhusudhan2014,mordasinivanboekel2016,cridlandpudritz2016}. The general idea is to compare a planet's C/O to the C/O predicted for the disk's solid and gas phases, and using the planet's bulk enrichment to determine whether the planet's metal enrichment is dominated by solid or gas accretion. From this it may be possible to infer where in the disk a planet formed. If this formation location were to be conclusively shown to be further away from the star than the planet's current location, this would be a proof for orbital migration. In what follows we attempt such an analysis using the C/O ratio inferred for HR~8799e. We neglect the, potentially very important, effect of compositional gradients in the planet, that may lead to atmospheric abundances being different from the bulk of the planet \citep[e.g.,][]{lecontechabrier2012,vazanhelled2018}.

The C/O value that has been reported for the host star HR~8799 is ${\rm C/O}_{\rm star}=0.56$ \citep{sadakane2006}. The author finds that C and O have roughly solar abundances. The star is a $\lambda$-Bo\"otis-type star, meaning that its iron peak elements are subsolar ($\rm [Fe/H]$ measured for iron specifically is $-0.55\pm 0.1$, \citealt{sadakane2006}). Because the analysis presented in this sub-section hinges on C/O and the total metal content of the planet, we neglect this additional information for now. We note, however, that if the composition of the  photosphere of HR~8799 is representative for the disk from which its planets formed, assuming  solar abundance ratios for all elements except O\footnote{We vary C/O by varying O.} during the atmospheric modeling is problematic.

To constrain HR~8799e's formation history, we make the assumption that the disk from which the HR~8799 planets formed had a composition as specified in Table 1 of \citet{oebergwordsworth2019}. The authors of the paper present this as a model for the young solar nebula. Because the C/O ratio of HR8799 \rch{(${\rm C/O} = 0.56$)} is \rch{essentially} solar \rch{(${\rm C/O} = 0.55$, see \citealt{asplund2009})} and we are most interested in the molecular volatiles, which represent the largest mass reservoir for solid planetary building blocks, we deem this assumption acceptable, even though HR8799 is a $\lambda$-Bo\"otis-type star.

Using the planetary mass of $4.81_{-3.33}^{+8.78}$~$\mj$, the metallicity of $0.48_{-0.29}^{+0.25}$ and the C/O of $0.60_{-0.08}^{+0.07}$, inferred from our spectral retrieval with \ptrad, we find that HR~8799e is likely heavily enriched in ices, accreted in forms of pebbles and planetesimals, and most likely formed outside of the CO iceline. This conclusion was obtained from fitting the O/H and C/H content of HR~8799e, derived from our spectral retrieval, with an abundance model of a planet that forms in a disk as defined in \citet{oebergwordsworth2019}. For this we treated the planets mass, accreted solid mass, and the accretion locations of the solids and gas as free parameters. For the planet mass a prior based on our spectrally retrieved mass was assumed. The planetary O/H and C/H was then fitted with \emph{PyMultiNest}. Specifically, due to the increased planetary metallicity of $0.48_{-0.29}^{+0.25}$, we find that the planet has accreted between 65 and 360~$\mearth$ of ices (1-$\sigma$ range) that are mixed into its envelope and atmosphere. The large uncertainty stems from the large mass and metallicity uncertainties from our spectral retrieval. The atmospheric metal content of HR~8799e appears to be dominated by solids due to the large inferred atmospheric enrichment \citep[e.g.,][]{espinozafortney2017}. If the metal content were dominated by gas accretion the atmospheric metallicity is expected to be close to, \rch{or smaller than}, stellar. As we find that the planet has a C/O ratio consistent with its host star (we find ${\rm C/O}=0.60_{-0.08}^{+0.07}$ for HR8799e) this could mean that the planet has formed outside of the CO iceline. In particular, we derive that a formation location outside the CO iceline is more than twice as likely compared to a formation inside the CO iceline (more details on this analysis will be published in an upcoming study). This is explained by the fact that only outside the CO iceline the solid material in the disk will have the stellar value, see \citet{oeberg2011}. \rch{In contrast, C/O values of 0.3, which is the lower boundary of the atmospheric grid approached in our \emph{Exo-REM} retrieval, are possible if the planet formed within the CO iceline, where \ce{H2O} and \ce{CO2} dominate the ice composition. If HR~8799e did form outside the CO iceline, this could mean that all HR~8799 planets formed outside the CO iceline, because HR8799e is the innermost planet of the HR8799 system.} This will have to be tested by deriving C/Os and metallicities for all HR~8799 planets. The C/O analysis of the HR~8799 planets by \citet{laviemendoca2017} is consistent with this assessment: for the HR~8799b and c, where the authors succeeded in deriving C/O values, they found ${\rm C/O} \geq {\rm C/O}_{\rm star}$.

If HR~8799e truly formed outside the CO iceline, this also allows to put constraints on the planet's possible migration. In \citet{oebergwordsworth2019}, the CO iceline is situated at $\sim$20~au for the young solar nebula. In the disk around HR~8799, the same temperature due to the irradiation of the star would have been reached at $\sim$45~au \citep{maroiszuckerman2010}, neglecting the evolution of the stellar luminosity at young ages. Because HR~8799e resides at $\sim 15$~au \citep{wanggraham2018}, this could imply that the planet migrated significantly. This would be consistent with the finding by \citet{wanggraham2018} that the HR~8799 planets needed to migrate in the gas disk after formation to get locked into a stable resonant orbit.

\rch{The model for the disk composition in \citet{oeberg2011,oebergwordsworth2019} is strongly simplified. The disk's properties such as temperature, surface density, and abundance profiles are assumed to be static. The iceline positions are determined from a simple thermodynamic stability analysis of the ice species. Processes such as the viscous and chemical evolution of the disk are neglected. However, chemical evolution and ionization of the disk material can be of crucial importance, as well as the initial composition of the disk at the start of the evolution, as shown by \citet{eistrupwalsh2016,eistrupwalsh2018}. Importantly, these studies describe how gas-grain chemistry may deplete CO from the gas phase within the CO iceline, condensing it in the form of CO$_2$, at the expense of also \ce{H2O}. Thus our conclusion regarding migration, based on the C/O ratio of HR~8799e, may be based on oversimplified disk chemistry assumptions.}

To assess the effect of properly treating the disk chemistry we used the ANDES physical-chemical code to compute a 2D steady-state disk physical structure and time-dependent chemistry for the HR~8799 disk \citep{2013ApJ...766....8A,2017ApJ...849..130M}.
\rch{The detailed setup of the model is described in Section \ref{sect:andes_setup}}. For this setup of the disk chemical model we found that the CO iceline lies at around 100~au, which would indicate that HR~8799e migrated even further after formation. However, due to the above-mentioned gas-grain chemistry, CO is converted into CO$_2$ ice effectively starting from around 20~au. This makes the solid C/O in the disk approach stellar values already at 20~au, such that HR~8799e may have formed as close as 20~au from the star. Hence realistic disk chemistry modeling could indicate that the planet migrated much less than when compared to simplistic disk abundance models.

\rch{Finally, we note that the nitrogen content may be a better way of constraining a planets formation location in the disk \citep{oebergwordsworth2019,bosmancridland2019}, where a large N-content corresponds to a formation in the outer parts of the disk. However, this would require to study planets cooler than HR~8799e, for which most of its accreted nitrogen is in the form of \ce{N2}, and therefore invisible due to the low \ce{N2} opacity.}

\subsection{Comparison of retrieval results with literature studies}
\label{sect:literature_comp}

Since their discovery \citep{maroismacintosh2008,maroiszuckerman2010,currieburrows2011} the HR8799 planets have been extensively studied. We provide a summary of the studies that exist on the HR8799 planets below and how they relate to our results for HR8799e. We start with the qualitative properties of the planet, before comparing our retrieved values with the ones reported by others.

\subsubsection*{Clouds}
In general, studies find that all HR8799 planets have comparable surface gravities and temperatures. In addition, all studies find that the HR8799 planets are dominated by thick clouds, as indicated by their red near-infrared (NIR) colors. For example, using NIR and MIR photometry and a grid of self-consistent atmospheric models, \citet{madhusudhanburrows2011} find that HR8799bcd are dominated by clouds much thicker than expected for field brown dwarfs. This is similar to an assessment already made by \citet{bowlerliu2010}, studying HR8799b. \citet{marleysaumon2012} came to the conclusion that thick clouds are required when studying HR8799bcd, but emphasized that this is not peculiar, and rather a consequence of cloud formation being gravity-dependent, a result that has also been borne out by the model calculations of \citet{charnaybezard2018}.

Considering the apparent lack of comparison brown dwarfs that resemble the HR8799 planets, it is important to remember that the earlier results in the literature are highly heterogeneous in terms of the available data (and models) that were used to arrive at a given conclusion. High quality spectra are important to truly unlock the planetary characteristics. For example, additional SPHERE spectroscopy obtained by \citet{zurlovigan2016} allowed \citet{bonnefoyzurlo2016} to show that especially HR8799de can be well fit with low-gravity cloudy brown dwarfs of the late L spectral type. For HR8799bc the picture is less clear, and \citet{bonnefoyzurlo2016} find that good comparison objects can only be identified when reddening T-dwarf spectra with iron or silicate extinction. Given the abundance of literature reporting on the cloudiness of the HR 8799 planets, \rch{our finding that HR 8799e is cloudy therefore does not come out of the blue (sky).}

\subsubsection*{Disequilibrium chemistry}
Disequilibrium chemistry has also been reported in the HR8799 planets by a variety of studies. For planets such as HR~8799bcde, disequilibrium chemistry means that CO is more and CH$_4$ is less abundant then predicted from chemical equilibrium, due to atmospheric mixing overruling the chemical reactions in the upper atmosphere. Using OSIRIS H and K band spectra at low resolution, \citet{barmanmacintosh2011} report on HR8799b exhibiting weaker \ce{CH4} absorption than expected, and that disequilibrium chemistry may be at play in order to decrease the \ce{CH4} abundance. Similar findings were also reported for the medium-resolution OSIRIS data for planets b and c, where \ce{CH4} was not detectable in c \citep{konopackybarman2013,barmankonopacky2015}. \citet{madhusudhanburrows2011} found that they had to neglect the 3.3~$\mu$m band containing too strong \ce{CH4} absorption when fitting their chemical equilibrium models to the HR8799bc data obtained by \citet{currieburrows2011} (there was only an upper limit available for d in the 3.3~$\mu$m band).
\citet{marleysaumon2012} reported disequilibrium chemistry to be important for HR8799bcd. \citet{skemermarley2014} report on disequilibrium chemistry for HR8799cd, based on NIR and MIR (narrow and broad band) photometry. Based on photometry, \citet{currieburrows2014} report the need for disequilibrium in planets b and c, but less strongly for d and e. \citet{laviemendoca2017} presented the first free retrieval analysis of the HR8799 planets and found that bc require disequilibrium chemistry, but not planets de. It is important to note, however, that no K-band spectroscopy was available for planet e, and that the K-band seemed heavily affected by systematics for planet d. Because the K band is important for detecting the presence of CO, and hence for detecting disequilibrium chemistry, their finding for d and e to be in chemical equilibrium has to be taken with caution. Using SPHERE and the same GPI K band data as in \citet{laviemendoca2017}, \citet{bonnefoyzurlo2016} report that HR 8799de can be fit well with chemical equilibrium models when using a non-scattering-cloud Exo-REM \citep{baudinobezard2015} grid. With the new Exo-REM models by \citet{charnaybezard2018}, which include chemical disequilibrium and scattering clouds, it is found that HR8799e requires disequilibrium to explain the GRAVITY K band data reported in \citet{lacournowak2019}. In summary, while the consensus in the literature regarding HR8799e is not entirely clear, we corroborate the finding of  \citet{lacournowak2019}, which is based on a high-S/N spectrum in the K-band, that HR8799e is affected by disequilibrium chemistry.

\begin{table*}[t!]
\centering
{ \footnotesize
\begin{tabular}{l|lllllll}
\hline \hline
Parameter (unit) & This study & This study & \rch{This study} &  B16 & L17 & G18 & C14 \\
& \ptrad & \emph{Exo-REM2} & \rch{\emph{Exo-REM2}} & \emph{Exo-REM1}$^{\rm (a)}$ & \emph{Helios-r} & \emph{PHOENIX}$^{\rm (a)}$ & \emph{COOLTLUSTY} \\
& & & \rch{fix SPHERE/} & & & & \\
& & & \rch{GPI scaling} & & & & \\ \hline
${\rm log}(g)$ (cgs) & $4.00_{-0.52}^{+0.46}$ & \rch{$3.82_{-0.13}^{+0.26}$} & \rch{$4.23_{-0.03}^{+0.04}$} & 3.7 & $3.9_{-0.05}^{+0.05}$$^{\rm (d)}$ & 3.5 & 4 \\
$T_{\rm eff}$ (K) & $1154_{-48}^{+49}$ & \rch{$1071_{-50}^{+61}$} & \rch{$1180_{-17}^{+16}$} & 1200 & --$^{\rm (e)}$ & 1100 & 1000 \\
$R_{\rm P}$ ($\rj$) & $1.12_{-0.09}^{+0.09}$ & \rch{$1.32_{-0.17}^{+0.15}$} & \rch{$1.08_{-0.02}^{+0.03}$} & 1.0 & $1.2_{-0.1}^{+0.05}$$^{\rm (d)}$ & 1.3 & $R_{\rm evo}^{\rm (h)}$\\
${\rm C/O}$ & $0.60_{-0.08}^{+0.07}$ & {$< 0.3^{\rm (b)}$} & \rch{$< 0.3^{\rm (b)}$} & -- & $\rightarrow 0^{\rm (f)}$ & -- & -- \\
${\rm [Fe/H]}$ & $0.48_{-0.29}^{+0.25}$ & {$> 0.5^{\rm(c)}$} & \rch{$> 0.5^{\rm(c)}$} & 0.5 & ~0.4$^{\rm (g)}$ & -- & -- \\ \hline
\end{tabular}
}
\caption{Comparison of reported properties of HR8799e, derived from spectral/photometric analyses. \rch{The second \emph{Exo-REM2} retrieval from our study fixed the SPHERE/GPI scaling factors to the best-fit values of \ptrad.} References: B16 \citep{bonnefoyzurlo2016}, L17 \citep{laviemendoca2017}, G18 \citep{greenbaumpueyo2018}, C14 \citep{currieburrows2014}. Code references: \emph{Exo-REM2} \citep[version of][]{charnaybezard2018}, \emph{Exo-REM1} \citep[version of][]{baudinobezard2015}, \emph{Helios-r} \citep[][]{laviemendoca2017}, \emph{PHOENIX} \citep[models reported in][]{barmanmacintosh2011}, \emph{COOLTLUSTY} \citep[][]{sudarskyburrows2003,hubenyburrows2003,burrowssudarsdy2006}. Notes: (a) these publications compared their observations to more than one model, we report the best-fit values of the models that provide the best fit to the data, either as stated by the authors or by visual inspection. (b) the best-fit model was trending into the boundaries of the Exo-REM grid, and only a boundary value can be reported for C/O. (c) same as (b), but for the [Fe/H] parameter. (d) as read of by eye from their corner plot. (e) not specified. (f) no K-band spectrum was available for their analysis, such that the CO abundance could not be constrained. Their retrieved C/O ratio is pushed against the 0 boundary. (g) as derived from their stated O/H ratio, in comparison to the solar O/H ratio. (h) consistent with radii derived from evolutionary models.}
\label{tab:hr8799e_comp}
\end{table*}

\subsubsection*{Reported modeling challenges}
When comparing models to observations of the HR~8799 planets a few problems have been identified. First, the models can converge toward high effective temperatures and thus small radii, in order to conserve the total flux of the planet within the model. Some radii that have been reported are smaller than expected from theoretical standpoints. Due to electron degeneracy pressure, the radii of gas giant planets and brown dwarfs are always above 0.75~$\rj$ and even this lowest limit is only reached after 10~Gyr of contraction, for brown dwarfs of 70~$\mj$ \citep{chabrierbaraffe2009}. For ages up to 5~Gyr, radii are above 1~$\rj$ for masses up to 30~$\mj$ \citep{mordasinialibert2012}. For objects with ages below 100~Myr, the minimum radius is therefore expected to be above 1~$\rj$ \citep{marleysaumon2012}. A radius that is too small when compared to these theoretical constraints hints at shortcomings in the atmospheric model being used. Examples for such small inferred radii are found in \citet{barmanmacintosh2011} (who report $R=0.75$~$\rj$ for HR8799b), \citet{greenbaumpueyo2018} (who report that some, but not all, of the model grids they tried have radii below 1~$\rj$ for HR8799cde) and \citet{bonnefoyzurlo2016} (who reported the same for some but not for all of the models they applied to HR8799bcde). One approach is to reject such unphysical radii right from the start, by tying the spectral model fitting to evolutionary models, which guarantees that physically consistent radii are used for the spectral analysis  \citep{marleysaumon2012}. Because this can worsen the spectral fit and thus lead to additional biases concerning inferred properties of the atmosphere, it is questionable how much is gained from such an approach \citep{barmankonopacky2015}. In this regard, limiting the radius in our HR8799e retrieval to a minimum value of 0.9~$\rj$ can be seen as a compromise between these two approaches, and our best-fit radius of $1.12_{-0.09}
^{0.09}$~$\rj$ is above the 1~$\rj$ limit described above.

In a similar vein, it is useful to check whether the retrieved values of the planet's effective temperature, surface gravity and radius (hence also its luminosity and mass) are consistent with evolutionary models. Considering the evolutionary plots presented in \citet{marleysaumon2012} (their figures 8 and 11) and following their analysis, places our median log$(g)$ and $T_{\rm eff}$ values between their 10 and 30~Myr isochrones, although our log$(g)$ uncertainties also allow for ages in excess of 100~Myr. These values are consistent with the ages that can be inferred for the HR8799 system \citep[between 30 and 60 Myr, see][for a more complete discussion]{marleysaumon2012}. Similarly, our median log$(g)$ and $T_{\rm eff}$ values lie between the evolutionary tracks of 5 and 10~$\mj$ models, and again our large uncertainties on log$(g)$ allow for masses well below 5~$\mj$ and in excess of 10~$\mj$. The mass we derive from HR8799e's spectrum, $M_{\rm P}=4.81_{-3.33}^{+8.78}\mj$ is certainly consistent with this assessment. We note here that \citet{marleysaumon2012} made the assumption of a hot start evolution in their work, and that the HR8799 system is young enough for hot and cold start differences to play a role. It was found by \citet{spiegelburrows2012,marleysaumon2012,marleaucumming2014}, however, that at least a classical cold start assumption \citep{marleyfortney2007} is ruled out for these planets. In any case, recent theoretical modeling of the physics of the accretion shock \citep{mkkm17,mmk19} and of the structure of accreting planets \citep{berardo17,berardocumming17,cumming18} suggests that hot starts are more likely. The spectroscopic mass derived in our study is also consistent with models studying the orbital evolution of the HR~8799 system, where it was found that the mass of HR~8799e has to be below 7.6~$\mj$ to ensure orbital stability \citep{wanggraham2018}.

In addition, models that assume a homogeneous cloud cover have been reported to have trouble at reproducing all the MIR photometry simultaneously, especially the 3.3~$\mu$m, L' and [4.05] bands.
The use of patchy cloud models, or models mixing clouds of different vertical extent, has been shown to be one promising way of solving this problem \citep[see, e.g.,][]{currieburrows2011,skemerhinz2012,currieburrows2014,skemermarley2014}. With this in mind it is interesting to see that our high likelihood retrieval models reproduce both the 3.3~$\mu$m and L' band photometry, {\it without including these data points in the retrieval}. This would speak against a heterogeneous cloud cover being necessary to explain the data. A similar result was found by \citet{bonnefoyzurlo2016} when fitting the HR~8799 planets with the Exo-REM code \citep{baudinobezard2015} in the non-scattering, chemical equilibrium version. It is important to note, however, that neither \citet{bonnefoyzurlo2016} nor we can reproduce the [4.05] band photometry of \citet{currieburrows2014}, which is consistently higher than our best-fit spectra, and all the best-fit models presented in \citet{bonnefoyzurlo2016}, for all of the HR8799 planets. Thus, a heterogeneous cloud cover cannot be ruled out. Given the prevalence of variability of, especially, L-T dwarf transition objects \citep[see, e.g.,][]{apairadigan2013,crossfield2014}, such a heterogeneous cloud cover could be expected for the HR~8799 planets.

\subsubsection*{Quantitative comparison to the literature}
Finally, we compare our retrieved parameter values for HR8799e with those reported in the literature. All values are listed in Table \ref{tab:hr8799e_comp}. Our derived gravity value, ${\rm log}(g) = 4.00_{-0.52}^{+0.46}$, falls within the values reported in the literature, which range from 3.5 to 4. The same holds for the effective temperature, for which we retrieve $1154_{-48}^{+49}$~K. This value is bracketed by the reported values, ranging between 1000 and 1200 K. Similarly, our retrieved radius value  ($1.12_{-0.09}^{+0.09} \ \rj$) falls between the reported values of 1 to 1.3~$\rj$. Including ours, four out of six data--model comparison studies also varied the metallicity. Reported values are between 0.4 and 0.5, and our value (${\rm [Fe/H]} = 0.48_{-0.29}^{+0.25}$) is consistent with their assessments. C/O  deserves a more detailed discussion. Only one other study considered the atmospheric C/O. The retrieved values range from 0 to $0.60_{-0.08}^{+0.07}$, the latter being the value we derive with \ptrad \ in this study. As discussed in Section \ref{sect:exo_rem_fit}, the Exo-REM fit runs into the grid boundaries for both C/O and [Fe/H], making it difficult for us to assess whether this trend to low C/O ratios is actually merited by the data \rch{or whether this parameter is used to copmensate for the too stringent cloud description}. The value reported by \citet{laviemendoca2017} (${\rm C/O}=0$) suffers from the fact that K-band spectroscopy was not available at the time of their study, such that the CO abundance could not be determined. They only report an upper limit for the atmospheric C/H value, and their best-fit spectrum does not show any CO absorption in the K-band, which we detect in the GRAVITY data. Given these caveats we conclude that a comparison to the C/O ratios derived in other studies is at this point inconclusive.

\section{Summary}
\label{sect:summary}

We present a new version of our retrieval radiative transfer code \ptrad \ \citep{mollierewardenier2019} to which we added the effect of multiple scattering. This enables us to run free retrievals on cloudy self-luminous objects such as directly imaged planets and brown dwarfs. This updated version of \ptrad \ will be available on the \ptrad \ website soon\footnote{\url{https://petitradtrans.readthedocs.io}} and is already available now, upon request.

Running verification retrievals on synthetic observations, we found that we can retrieve the input parameters. The high dimensionality of the input model can lead to small offsets within the observational uncertainties, however. Increasing the number of live points in our retrievals with the nested sampling method improves this, but we expect this to be a persisting problem for models with a large number of free parameters, especially for high S/N observations which lead to narrow posterior distributions \rch{and thus let a smaller fraction of the prior volume be of interest}.

We tested two different cloud models.  The first is the physically motivated \citet{ackermanmarley2001} cloud model. Our second cloud model simply retrieves the wavelength properties of the cloud opacities. When running retrievals with Cloud Model 2 on mock observations made with Cloud model 1, we find that we can get an excellent fit but with biased atmospheric parameters. In particular the planet's photosphere is found to be more isothermal and less cloudy than the input, which mimics the shallow temperature gradients predicted by  \citet{tremblinamundsen2015,tremblinamundsen2016,tremblinchabrier2017,tremblinpadioleau2019}. Thus, retrievals alone will likely not be enough to investigate whether shallow temperature gradients indeed occur in such thought-to-be-very-cloudy atmospheres. Retrieval analyses could be aided by longer wavelength data in the mid-IR, however, which could reveal the spectral features of cloud particles at $\sim$10~micron \citep{cushingroellig2006}.

We ran our retrieval setup on archival GPI, SPHERE, and partially new GRAVITY data for the directly imaged planet HR~8799e, using Cloud Model 1. Applying such data-driven, free retrievals for directly imaged planets becomes possible with our high S/N observations, especially GRAVITY's K band spectra at a spectral resolution of $R = 500$. In addition, observations in the K band can probe \ce{H2O}, \ce{CH4} and \ce{CO} features and are therefore crucial for constraining atmospheric disequilibrium chemistry and the atmospheric C/O.

We are able to fit the observations well, and confirm a cloudy atmosphere dominated by disequilibrium chemistry. We also compare our retrieved atmospheric spectra to archival photometric observations in the L', 3.3~$\mu$m, [4.05] and M band. Our spectra are consistent with all points, except for the [4.05] band point, although the photometry has not been included in the retrieval. The L'$-$3.3~$\mu$m and L'$-$[4.05] band colors have been suggested to require heterogeneous cloud coverage to explain the data, and we cannot confirm this for the L'$-$3.3~$\mu$m color.

The posterior parameter values we retrieve for the atmospheric ${\rm log}(g)$, $T_{\rm eff}$ and [Fe/H] are consistent with previous studies, and hot start evolutionary calculations. For the first time, we successfully constrain the C/O of HR~8799e and find that it is $0.60_{-0.08}^{+0.07}$, which is consistent with stellar. {\it Running additional retrievals on HR~8799e with Cloud Model 2, we find that the retrieved planetary C/O, [Fe/H] and ${\rm log}(g)$ are identical to the values found with the nominal Cloud Model 1, therefore independent of our cloud model choice. This is noteworthy as the Cloud Model 2 retrievals lead to less cloudy, more isothermal atmospheres. This indicates that C/O may be a quite robust outcome of the retrievals.} We also fit the HR~8799e spectrum with the self-consistent code \emph{Exo-REM}, which uses a state-of-the-art one-dimensional cloud model, scattering, and disequilibrium chemistry. Our free retrieval results compare well to the \emph{Exo-REM} fit, except for the C/O ratio. With \emph{Exo-REM} the C/O ratio is driven into the grid boundary (${\rm C/O}< 0.3$), as is the metallicity derived from \emph{Exo-REM} (${\rm [Fe/H]}> 0.5$). This makes the comparison between the free retrieval with \ptrad \ and \emph{Exo-REM} difficult for C/O. A larger \emph{Exo-REM} grid, which also varies the free parameters of its cloud prescription, may resolve this issue.

Using our retrieved C/O and metallicity, \rch{and a highly simplified disk model}, we find that HR~8799e could have formed outside the CO iceline. \rch{This would imply that the planet migrated significantly.} Because HR8799e is the innermost planet of the HR8799 system, this could indicate that all HR8799 planets formed outside of the CO iceline. Similar formation distances, \rch{relative to the icelines}, have been theorized for Jupiter in the Solar System \citep{oebergwordsworth2019,bosmancridland2019}. Using sophisticated gas-grain chemical modeling for the protoplanetary disk we find that the planet could also have formed more closely to the star, but outside the \ce{CO2} iceline. This would require less migration.

\section{Outlook}
Here we introduced the first version of our retrieval framework for cloudy scattering atmospheres. With this we introduce a versatile tool for interpreting the spectra of directly imaged planets and brown dwarfs. At the same time, it is clear that there are many avenues for improving and testing our method.

One could explore different pressure-temperature parameterizations, cloud model setups, or abundance models. The latter could mean, for example, retrieving absorber abundances independently, as is often done in retrieval studies. Alternatively, the assumption of chemistry could be kept, while retrieving not just C/O (which we varied by changing O) and metallicity, but by retrieving C/H, O/H, and other atomic abundance ratios instead \citep[as was done in, e.g.,][]{laviemendoca2017,spakesing2019}. Another important addition will be to include a parameterization for heterogeneous  cloud coverage.

It is also crucial to test what happens when retrieving atmospheric parameters using a model setup that is different from the one used to generate a synthetic observation. In this way one can begin to quantify the uncertainties and biases of retrieved parameters given the model choices, and how robust certain parameters are against using a wrong model. An example is the robustness of C/O that we found in our results here, when using different cloud models. In principle, the most likely among a set of models can be found using the Bayes factor, computed with the model evidences derived from nested sampling. However, a given model may be worse than another in terms of model assumptions, which could be highly unphysical, while still being favored by a Bayes factor analysis. Thus, such comparison retrieval studies offer additional insight regarding the real parameter uncertainties, including the modeling choices.

On the observational side, additional data in the mid-IR, ideally spectroscopy, is necessary to explore the properties of clouds further. This is because the L', 3.3~$\mu$m and [4.05] band may encode information about a heterogeneous cloud cover. Excitingly, the outer planets (HR~8799bcd) will be studied in the mid-IR, using NIRSpec IFU spectroscopy with the JWST\footnote{\url{http://www.stsci.edu/jwst/phase2-public/1188.pdf}}. HR~8799bcde will also be studied with photometry in the mid-IR with JWST \footnote{\url{http://www.stsci.edu/jwst/phase2-public/1194.pdf}}. ESO's ERIS\footnote{\url{http://www.eso.org/sci/facilities/develop/instruments/eris.html}} instrument, to be mounted at the VLT, is a promising option, as well as KPIC, to be mounted on Keck II \citep{mawetbond2018}. ESO's imminent CRIRES+ instrument may offer the possibility to study the HR~8799 planets at high spectral resolution, where the S/N of the planetary flux measurement could be boosted using the cross-correlation method \citep[e.g.,][]{hoeijmakersschwarz2018}, which also allows for carrying out retrievals \citep{brogiline2018}. Further in the future, the METIS spectrograph \citep{brandlfeldt2014} of ESO's upcoming ELT telescope as well as the PSI instrument \citep{skemerstelton2018} on the TMT will be excellent instruments for mid-IR observations.
For studying the potential absorption feature of silicate clouds at 10~micron, JWST will again be an excellent instrument. Here one approach could be to obtain mid-IR spectra of HR8799 planet analogs such as PSO~J318\footnote{\url{https://www.stsci.edu/jwst/phase2-public/1275.pdf}}.

Lastly, also constraining basic parameters of the HR~8799 planets better could be of great help. If astrometry were to give a mass estimate for the HR~8799 planets, a prior on the planet mass would lead to a better ${\rm log}(g)$ and therefore [Fe/H] inference during the retrievals (these two parameters being correlated). This in turn would also help to understand the planet's formation better, because the planetary metallicity can be regarded as a measure for the relative importance of the solid body accretion of a planet.

\begin{acknowledgements}
\rch{We would like to thank Joanna Barstow for a thorough referee report, which greatly improved the quality of this paper. We also thank the A{\&}A editor, Emmanuel Lellouch, for additional comments.} P.M. thanks M. Line, J. Zalesky, and M. Min for insightful discussions. P.M. acknowledges support from the European Research Council under the European Union's Horizon 2020 research and innovation program under grant agreement No. 832428. T.S. acknowledges the support from the ETH Zurich Postdoctoral Fellowship Program. G.-D.M. acknowledges the support of the DFG priority program SPP~1992 ``Exploring the Diversity of Extrasolar Planets'' (KU~2849/7-1) and from the Swiss National Science Foundation under grant BSSGI0\_155816 ``PlanetsInTime''. Part of this work has been carried out within the framework of the National Centre of Competence in Research PlanetS supported by the Swiss National Science Foundation.
A.V. and. G. O. acknowledge funding from the European Research Council (ERC) under the European Union's Horizon 2020 research and innovation programme (grant agreement No.~757561). I.S. acknowledges funding from the European Research Council (ERC) under the European Union's Horizon 2020 research and innovation program under grant agreement No 694513. P.G. was supported by Funda\c{c}\~{a}o para a Ci\^{e}ncia e a Tecnologia, with grants reference UIDB/00099/2020 and SFRH/BSAB/142940/2018. T.M. acknowledges support by the grant from the Government of the Russian Federation 075-15-2019-1875, ``Study of stars with exoplanets''. D.S. acknowledges support by the Deutsche Forschungsgemeinschaft through SPP 1833: ``Building a Habitable Earth'' (SE 1962/6-1). A.Z. acknowledges support from the FONDECYT Iniciaci\'on en investigaci\'on project number 11190837. R.G.L. acknowledge support by Science Foundation Ireland under Grant No. 18/SIRG/559. This work benefited from the 2019 Exoplanet Summer Program in the Other Worlds Laboratory (OWL) at the University of California, Santa Cruz, a program funded by the Heising-Simons Foundation.
\end{acknowledgements}

\bibliographystyle{aa}
\bibliography{mybib}{}

\appendix

\section{K-table mixing}
\label{sect:ktab_mix}
This section describes our process for obtaining the total cumulative opacity distribution function, hereafter called k-table, of the atmosphere. The resulting k-table will contain the contribution of all absorbers. This combined k-table is required when including scattering during the computation of emission spectra. The combination is achieved by sampling the k-tables of the individual absorber species.

\subsection{Gaussian quadrature grid definition}
For every species, \ptrad \ stores k-tables as a function of pressure and temperature. The spectral bins for which the k-tables are stored have the width $\Delta \lambda$. The width varies as a function of wavelength, because it is chosen such that $\lambda/\Delta\lambda=1000$. Within these bins the opacity of each species is stored as a function of the cumulative probability $g$, where $g=0$ denote the lowest opacity values, and $g=1$ the highest values. These are the k-tables.
\rch{An introduction to correlated-k, and why 
such a k-table treatment is useful to speed up calculations when compared to line-by-line calculations, is given in \citet{irwinteanby2008}, their Section 2. Especially their Figure 1 illustrates the definition of $g$ as the spectral coordinate, when compared to the wavelength.} 
At low pressures, where the effect of pressure broadening is small, the k-tables of a line absorber will have a low-opacity tail extending over most of the $g$ values. In addition, the line cores will give rise to a steep increase in opacity (by orders of magnitude), when approaching $g$ values of unity. In order to sufficiently resolve the cumulative opacity distribution for low pressures, where the rise to highest opacity values will occur over a very narrow range at high $g$ values, we split our $g$ grid in two parts. The first is a Gaussian quadrature grid with coordinate $g^{\rm low}$ extending from $g^{\rm low}=0$ to $g^{\rm low}=0.9$, while the second is a Gaussian grid extending from $g^{\rm high}=0.9$ to $g^{\rm high}=1$. Both grids have eight points, and their weights $w$ have been rescaled such that
\beq
\sum_{i=1}^8 w^{\rm low}_i = 0.9 \ \ {\rm and} \ \ \sum_{i=1}^8 w^{\rm high}_i = 0.1.
\eeq
This weight rescaling guarantees that, for any $\lambda$- and $g$-dependent function $f$,
\begin{align}
\nonumber \left<f\right> &= \frac{1}{\Delta \lambda}\int_{\lambda-\Delta\lambda/2}^{\lambda+\Delta\lambda/2} f(\lambda)d\lambda \\
\nonumber &= \int_0^1 f(g)dg \\
&\approx \sum_{i=1}^8 \left[w^{\rm low}_i  f(g^{\rm low}_i)+ w^{\rm high}_i f(g^{\rm high}_i)\right].
\end{align}

\subsection{Sampling}
For obtaining the total k-table, we use the standard assumption for the on-the-fly combination of the opacities of different absorbers \citep[see, e.g.,][]{lacis_oinas1991,irwinteanby2008,mollierevanboekel2015,amundsentremblin2017}, namely that their opacity distribution functions are independent (also called {\it `random overlap'}). Making this assumption, an opacity value of the total k-table can be sampled by drawing opacity samples from the k-tables of each individual species, scaling them according to the respective abundances of the absorbers, and adding them.

By sampling the total k-table in this fashion many times, a good approximation of the total opacity distribution function can be constructed: the sampled values are simply sorted in magnitude, with the lowest value corresponding to $g=0$, and the highest value corresponding to $g=1$. This approximated k-table can then be interpolated back to the $g^{\rm low}$ and $g^{\rm high}$ values, to be ready for use in \ptrad.

\begin{figure}[t!]
\centering
\includegraphics[width=0.495\textwidth]{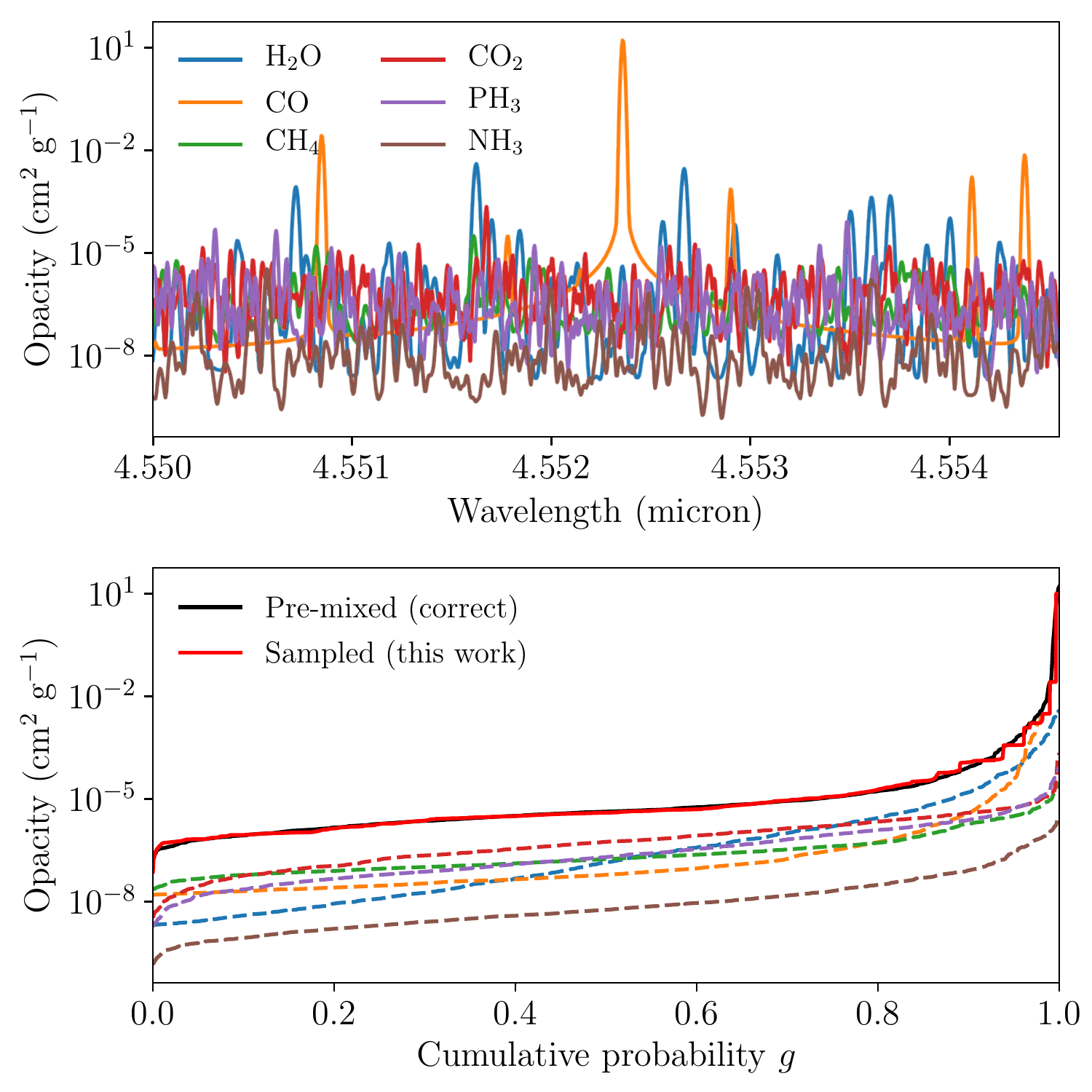}
\caption{{\it Upper panel:} wavelength-dependent opacities of \ce{H2O}, \ce{CO}, \ce{CH4}, \ce{CO2}, \ce{PH3}, and \ce{NH3}, in the spectral range between 4.55 and 4.5545 micron. {\it Lower panel}: k-table curves of the individual species (dashed lines, same color coding as in the upper panel), the total k-table obtained from premixing the opacities of all species in wavelength space (black solid line), and total k-table obtained with the method used in this work (red solid line).}
\label{fig:ck_sampling_demo}
\end{figure}

The sampling of a species' k-table could be done by drawing a uniformly distributed number between 0 and 1, and obtaining the opacity value by interpolating in the k-table of the species, to the so-drawn $g$ value. However, this step needs to be done a number of times, and a numerically less costly option is to randomly selected an index numbering the opacities of the species' 16-point k-table, and treating the tabulated opacity at this index as the sampled value. Doing this for all species, and scaling the opacities by the abundances, yields again a sample of the total k-table, when the samples are added.

One subtlety is that not every index is equally likely, because we store the opacities on two eight-point Gaussian quadrature grids for every species. Hence, when drawing samples $\kappa^{\rm tot}$ of the total opacity distribution function, the corresponding opacity value and its not-yet-normalized weight $w^{\rm tot}$ are
\beq
\kappa^{\rm tot} = \sum_{i=1}^{\rm N_{\rm spec}}X_i\kappa^i_{j(i)}, 
\eeq
\beq
w^{\rm tot} = \prod_{i=1}^{\rm N_{\rm spec}}w_{j(i)},
\eeq
where $X_i$ is the mass fraction of species $i$, and $j(i)$ denotes the k-table index sampled for species $i$.

The total sampled k-table can then be obtained by sorting the sampled $(\kappa^{\rm tot}, w^{\rm tot})$ pairs by their $\kappa^{\rm tot}$ values, and normalizing all sampled $w^{\rm tot}$ weights such that their sum equals unity. The corresponding $g$ coordinate of the sorted, sampled points is then equal to the cumulative sum of the rescaled $w^{\rm tot}$ weights. The thus-constructed k-table can then be used in \ptrad, after interpolating back to the $g^{\rm low}$ and $g^{\rm high}$ values.

In order to achieve a sufficient sampling of the opacity values in the $g^{\rm low}$ grid, we set up a sampling that draws the low indices three times as often.\footnote{This is done by randomly drawing the indices from an array containing (1,1,1,2,2,2,3,3,3,4,4,4,5,5,5,6,6,6,7,7,7,8,8,8,9,10,11,12,13,14,15,16).} To conserve their actual weight, we  multiply the weights $w^{\rm low}$ by a factor $1/3$ during this process.

Lastly, we speed up the k-table computation by neglecting those species $i$ for which
\beq
X_i\kappa_i(g=1) < 0.01 \times \max_{j}\left[X_j\kappa_j(g=0)\right].
\eeq
That is, a species whose maximum opacity value (at $g=1$) is more than 100 times smaller than the minimum opacity value (at $g=0$) that is the largest among all species, is deemed negligible.

\begin{figure}[t!]
\centering
\includegraphics[width=0.495\textwidth]{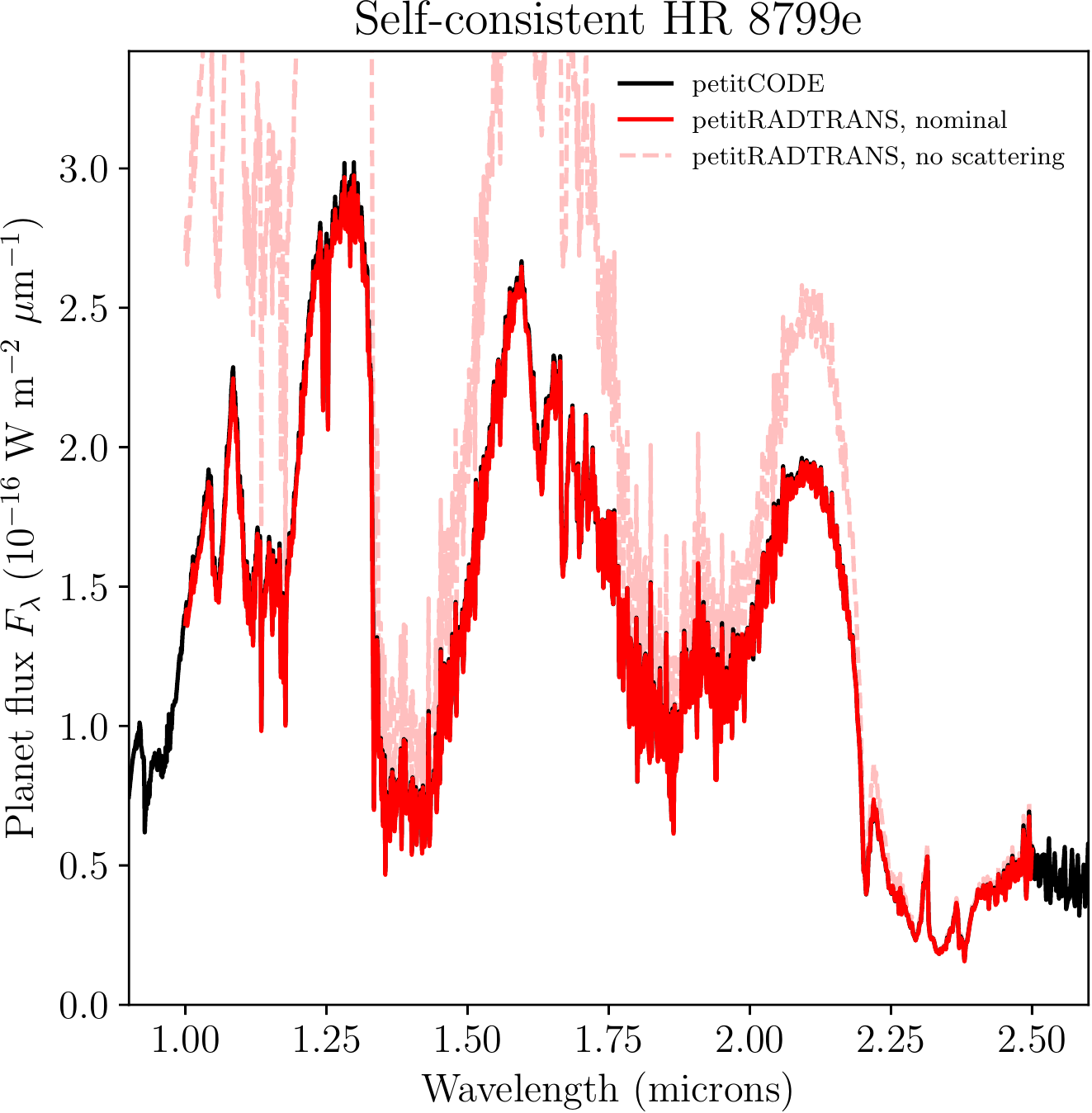}
\includegraphics[width=0.495\textwidth]{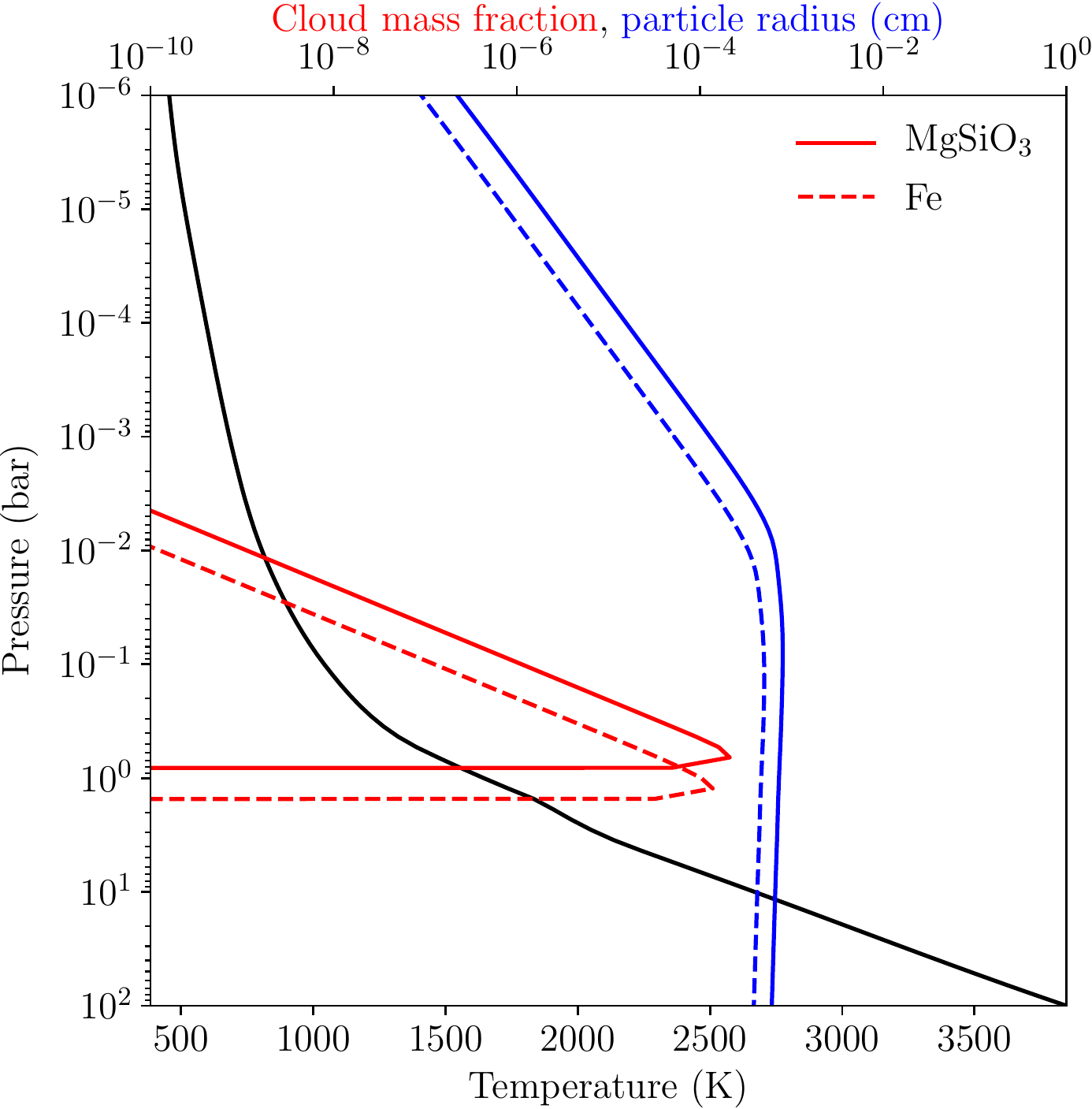}
\caption{{\it Upper panel:} emission spectrum of a synthetic, self-consistent model of HR~8799e. The \petit calculation is shown as a black solid line, whereas the \ptrad \ calculations, with and without scattering, are shown as red solid, and light red dashed lines, respectively. {\it Lower panel:} self-consistent atmospheric structure used for generating the spectra, showing the temperature (black solid line), cloud mass fractions (red lines), and cloud particle radii (blue lines) for \ce{MgSiO3} (solid lines) and Fe (dashed lines).}
\label{fig:spec_comp_paper}
\end{figure}

\subsection{Example of the k-table mixing}
In Figure \ref{fig:ck_sampling_demo}, we show an example of the k-table mixing outlined above. As in all spectral calculations presented in this work, each species is sampled 1,000 times. Here we used the HR8799e model calculated with \petit (see Section \ref{sect:ver_scat}) to show the k-table combination at a pressure of $10^{-5}$~bar, that is, at pressures so low that the opacities of each species will vary by multiple orders of magnitude in a given spectral bin. We reduced the abundances of \ce{PH3} and \ce{CH4} by factors of 0.001 and 0.01, respectively, to show an example of a k-table combination where no species is clearly dominating the opacity.

The upper panel of Figure \ref{fig:ck_sampling_demo} shows the wavelength-dependent opacities of \ce{H2O}, \ce{CO}, \ce{CH4}, \ce{CO2}, \ce{PH3}, and \ce{NH3}, in the spectral range between 4.55 and 4.5545 micron. This spectral width corresponds to the width of a typical \ptrad \ wavelength bin in the mid-infrared. The lower panel shows the k-table (i.e., $\kappa(g)$) curves of the individual species (dashed lines, same color coding as in the upper panel). It also shows the total k-table, obtained from premixing the opacities in wavelength space (black solid line), as well as the total k-table obtained with the method used in this work (red solid line), both of which agree very well.

\section{Verification of scattering treatment in cloudy emission spectra}
\label{sect:ver_scat}

Here we compare the radiative transfer results of \petit and \ptrad, when using a self-consistent atmospheric structure of \petit as the input for a \ptrad \ calculation. This is done to verify the k-table mixing (see Appendix \ref{sect:ktab_mix}) and scattering implementation (see Section \ref{sect:add_scat}) of \ptrad.

The model we use as an input is broadly motivated by the properties of HR~8799e, with the following input parameters: $T_{\rm eff}=1200 \ {\rm K}$, ${\rm log}(g)=3.75$ (with $g$ in ${\rm cm \ s^{-2}}$), ${\rm [Fe/H]}=0$, ${\rm C/O}=0.55$, $f_{\rm sed} = 3$, $\sigma_{\rm g} = 2$, $K_{zz}=10^{8.5} \ {\rm cm^2 s^{-1}}$, where $f_{\rm sed}$ is the settling parameter as defined in \citet{ackermanmarley2001}, $\sigma_{\rm g}$ is the width of the log-normal particle size distribution, and $K_{zz}$ is the atmospheric eddy diffusion coeffiecient. The rest of the symbols have their usual meaning. The clouds were assumed to consist of irregularly-shaped particles. For this we used cloud opacities calculated with the code by \citet{minhovenier2005}, which also makes uses of software by \citet{toonackerman1981}. The code assumes a distribution of hollow spheres (DHS) for the particles. Moreover, the condensates were assumed to be crystalline.

The upper panel of Figure \ref{fig:spec_comp_paper} shows the comparison of the resulting spectra of \petit and \ptrad, which agree excellently. In addition, we also show the \ptrad \ spectrum which would result from neglecting scattering. It is evident that scattering is an important process that needs to be accounted for. For completeness, the lower panel of Figure \ref{fig:spec_comp_paper} shows the self-consistent structures of the atmospheric temperature, cloud mass fraction and particle radius, resulting from the \petit calculation, which is used as an input for producing the spectrum with \ptrad.

\section{Verification retrieval with Cloud Model 1 and zoomed-in prior ranges}
\label{sect:app_tight_box}

\begin{table}[t!]
\centering
{ \footnotesize
\begin{tabular}{ll|ll}
\hline \hline
Parameter & Prior & Parameter & Prior \\ \hline
$T_1$ & $\mathcal{U}(0 , T_2)$ & ${\rm log}(\tilde{X}_{\rm Fe})$ & $\mathcal{U}(-1.3, 0)$ \\
$T_2$ & $\mathcal{U}(0 , T_3)$ & ${\rm log}(\tilde{X}_{\rm MgSiO_3})$ & $\mathcal{U}(-1.3, 0)$ \\
$T_3$ & $\mathcal{U}(0 , T_{\rm connect})$ & $f_{\rm sed}$ & $\mathcal{U}(1.8, 4.2)^{\rm (*)}$ \\
${\rm log}(\delta)$ & $P_{\rm phot} \in [10^{-3}, 100]$ & ${\rm log}(K_{\rm zz})$ & $\mathcal{U}(7.6, 9.9)^{\rm (*)}$ \\
$\alpha$ & $\mathcal{U}(1,2)$ & $\sigma_{\rm g}$ & $\mathcal{U}(1.2, 3)^{\rm (*)}$ \\
$T_0$ & $\mathcal{U}(500 , 2000 )$ & $R_{\rm P}$ & $\mathcal{U}(0.85, 1.05)^{\rm (*)}$ \\
$\rm C/O$ & $\mathcal{U}(0.45,0.6)^{\rm (*)}$ & ${\rm log}(g)$ & $\mathcal{U}(3.65, 4.2)^{\rm (*)}$ \\
$\rm [Fe/H]$ & $\mathcal{U}(-0.1,0.3)^{\rm (*)}$ & ${\rm log}(P_{\rm quench})$ & $\mathcal{U}(-6, -1)^{\rm (*)}$ \\ \hline
\end{tabular}
}
\caption{Priors of the ``zoomed-in'' verification retrieval of Cloud Model 1. Prior ranges marked with $^{\rm (*)}$ have been changed when compared to the initial run, so as to "zoom in" on the high likelihood region identified by the initial retrieval. The symbols have the same meaning as in Table \ref{tab:samp_prior}.}
\label{tab:zoom_in_verification_cloud_model_1}
\end{table}

\begin{figure*}[t!]
\centering
\includegraphics[width=0.95\textwidth]{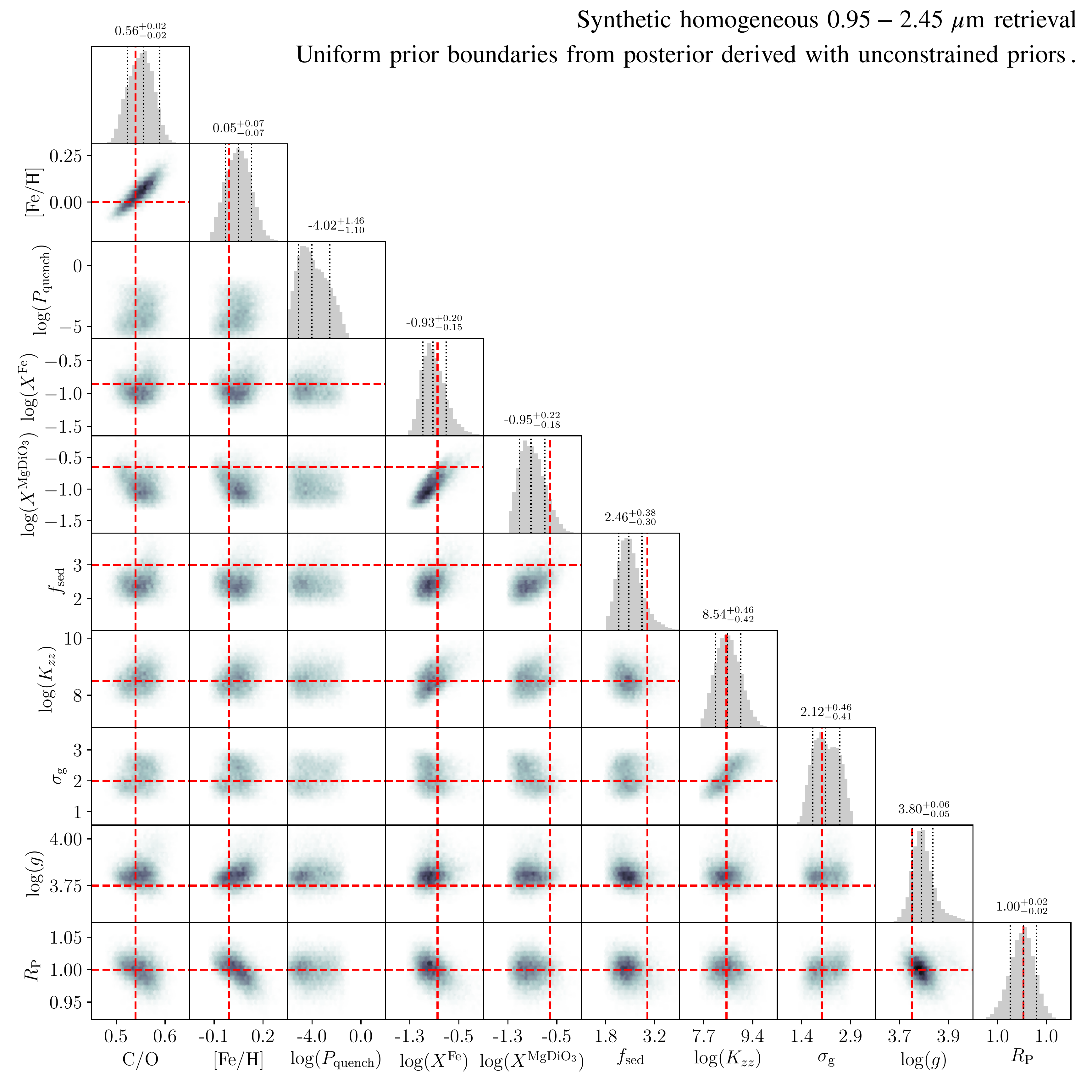}
\caption{Posterior distribution of the follow-up retrieval with Cloud Model 1 (see Section \ref{sect:ver_ret}). For this retrieval the uniform prior boundaries were chosen to enclose the regions of highest likelihood inferred in the original Cloud Model 1 retrieval, so as to test the effects of using the same number of live points in a smaller prior volume. The uniform prior boundaries are described in Appendix \ref{sect:app_tight_box}.}
\label{fig:corner_tight_box}
\end{figure*}

Figure \ref{fig:corner_tight_box} shows the follow-up retrieval of the verification retrieval with Cloud Model 1, discussed in Section  \ref{sect:ver_ret}. For this retrieval the uniform prior boundaries were chosen to enclose the regions of highest likelihood inferred in the original Cloud Model 1 retrieval, so as to test the effects of using the same number of live points in a smaller prior volume. The prior ranges are listed in Table \ref{tab:zoom_in_verification_cloud_model_1}.

\section{Retrieving Cloud Model 1 with Cloud Model 2}
\label{sect:ret_ver_test_cloud_model_2}
\begin{figure*}[t!]
\centering
\includegraphics[width=0.8\textwidth]{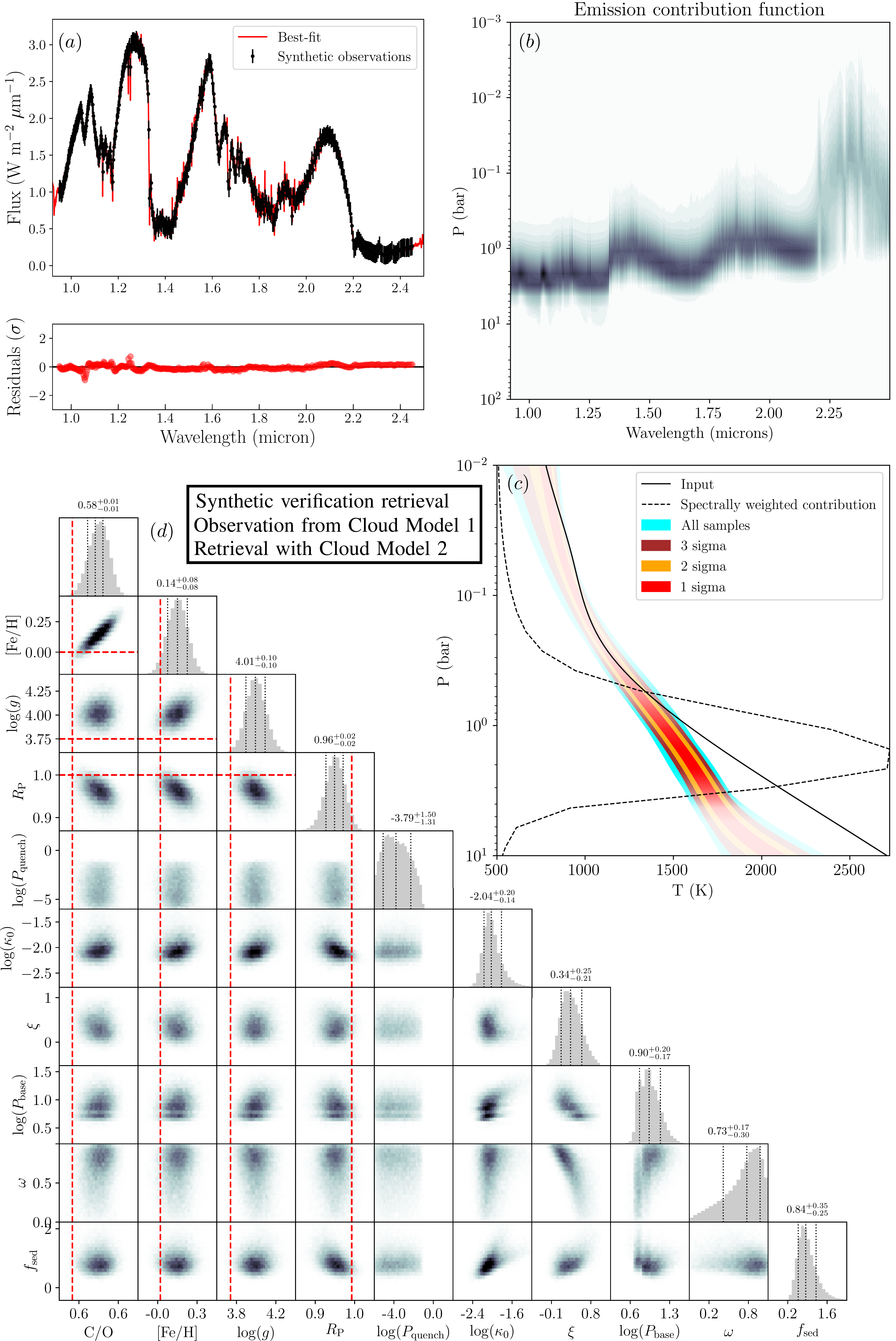}
\caption{Same as Figure \ref{fig:combine_all_alpha_12_1000_noise_free}, but showing the results when a synthetic spectrum made with Cloud Model 1 is retrieved using Cloud Model 2. This test is described further in Section \ref{sect:ret_CM1_with_CM2}.}
\label{fig:combine_all_alpha_12_simple_cloud}
\end{figure*}

In Figure \ref{fig:combine_all_alpha_12_simple_cloud} we show the result of the retrieval when a synthetic spectrum made with Cloud Model 1 is retrieved using Cloud Model 2. See Section \ref{sect:ret_CM1_with_CM2} for the discussion.

\section{Synthetic retrieval assuming HR8799e-like input data}
\label{sect:app_synth_hr8799e}
\begin{figure*}[t!]
\centering
\includegraphics[width=0.8\textwidth]{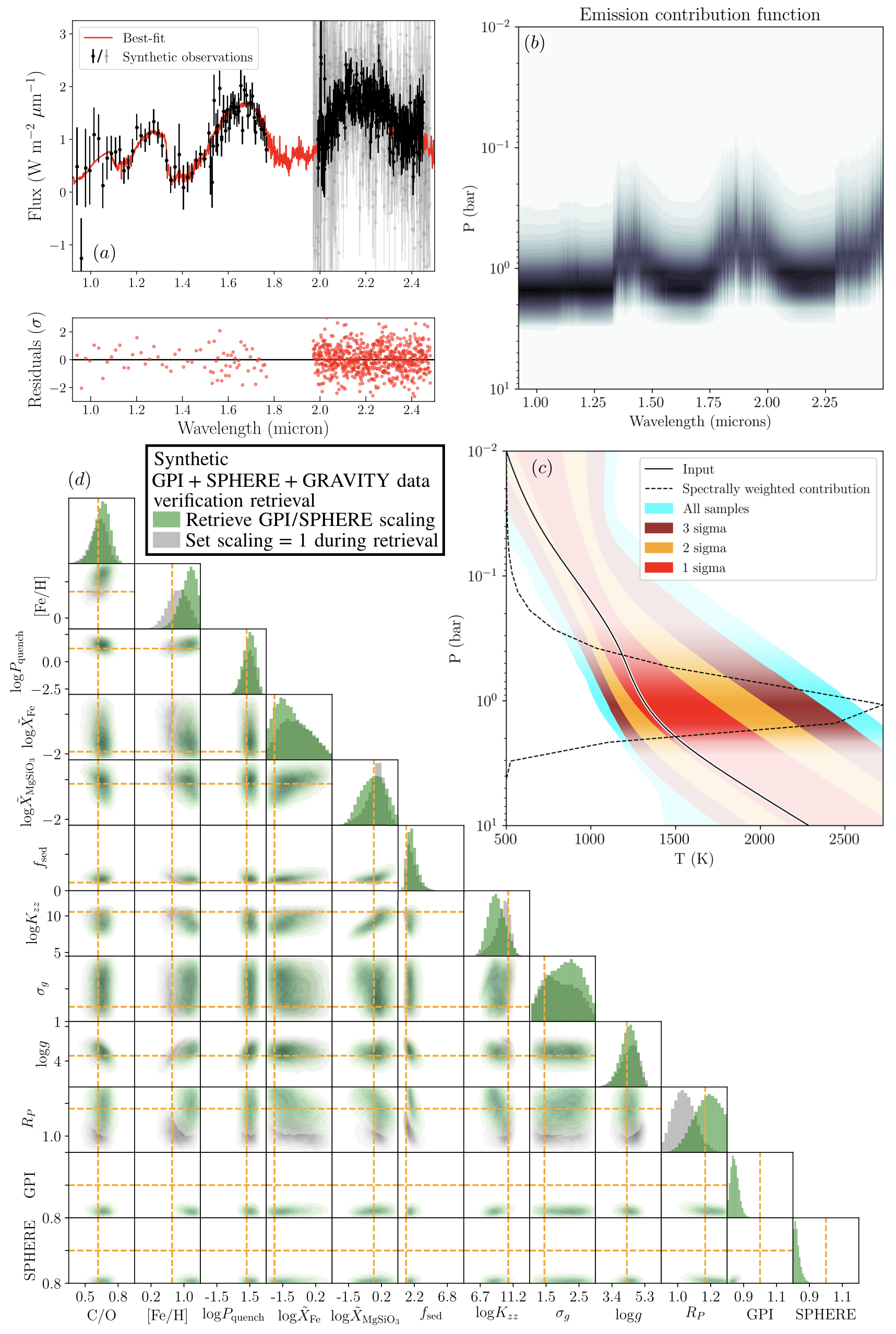}
\caption{Same as Figure \ref{fig:combine_all_alpha_12_1000_noise_free}, but showing the results when a synthetic spectrum of the same wavelength spacing and noise properties as the actual HR8799e data is retrieved. Green posteriors indicate the retrieval where the scaling parameters of the GPI and SPHERE spectra were retrieved, while gray posteriors indicate the results that neglected the scaling parameters. This test is described further in Section \ref{sect:test_hr8799e_model}. In Panel {\it (a)}, the black points in the K-band correspond to synthetic observations with the same data quality as the GRAVITY observations reported in \citet{lacournowak2019}, while the gray points show the data at the same quality as found for the two new GRAVITY observations presented in this work. All three synthetic GRAVITY data sets were fitted simultaneously.}
\label{fig:combine_all_ver_ret_like_real_data}
\end{figure*}

Figure \ref{fig:combine_all_ver_ret_like_real_data} shows the results of the synthetic retrieval using HR8799e-like data, described in Section \ref{sect:test_hr8799e_model}. The black points in the K-band correspond to synthetic observations with the same data quality as the GRAVITY observations reported in \citet{lacournowak2019}, while the gray points show the data at the same quality as found for the two new GRAVITY observations presented in this work. All three synthetic GRAVITY data sets were fitted simultaneously.

\section{Retrieval of HR8799e with the (non-nominal) Cloud Model 2}
\label{sect:app_hr8799e_CM2}
\begin{figure*}[t!]
\centering
\includegraphics[width=0.8\textwidth]{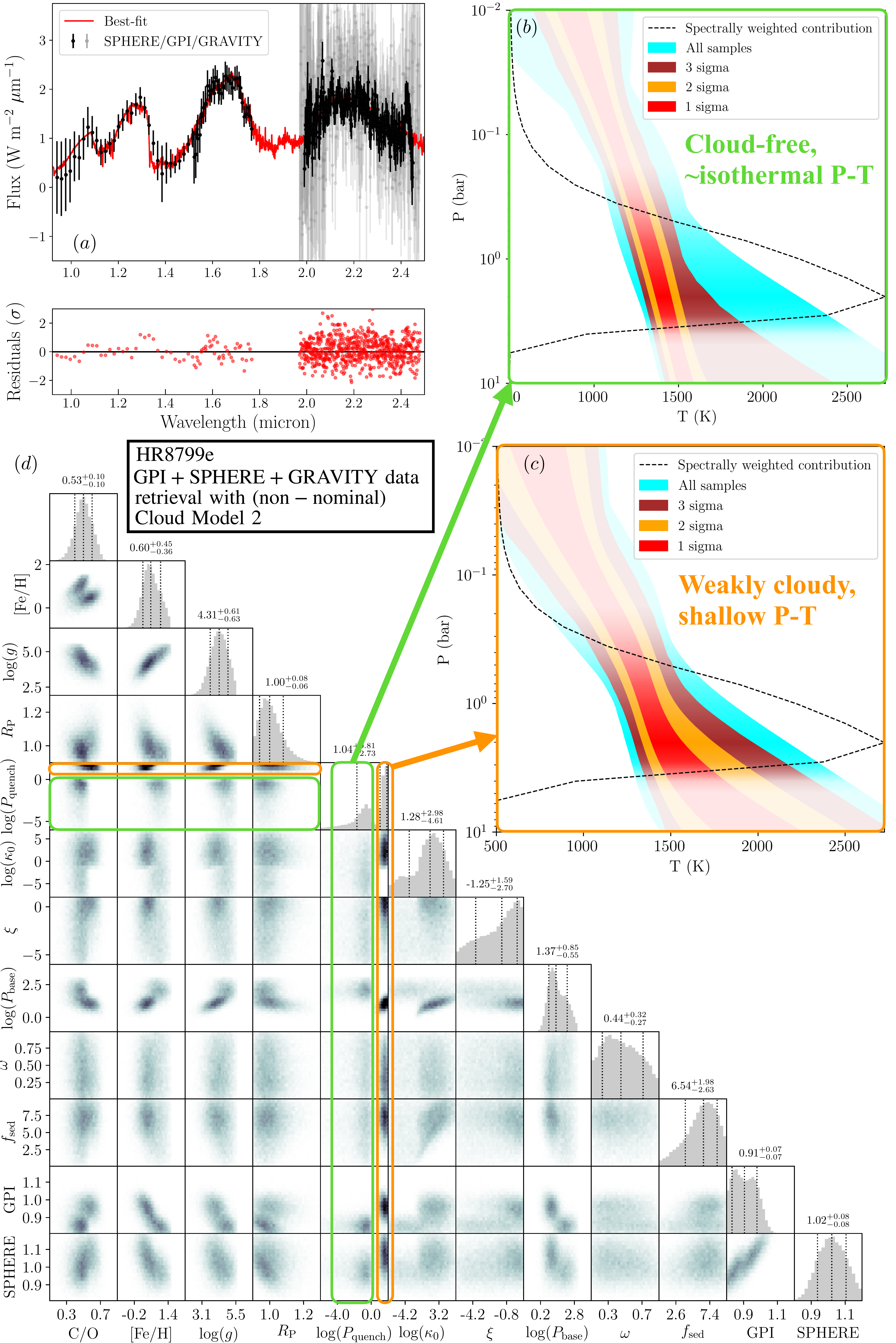}
\caption{Results of the non-nominal retrieval of HR8799e, using Cloud model 2. {\it Panel (a):} observations, best-fit spectrum and residuals. {\it Panel (b):} retrieved pressure-temperature confidence envelopes of the cloud-free isothermal solution of the retrieval. {\it Panel (c):} retrieved pressure-temperature confidence envelopes of the cloudy solution of the retrieval. The black dashed line shows the flux average of the emission contribution function of the global best-fit (clear and cloudy) from the posterior. {\it Panel (d):} 2-d posterior plot of the (non-nuisance) retrieved atmospheric parameters. In Panel {\it (a)}, the black points in the K-band correspond to the GRAVITY observations reported in \citet{lacournowak2019}, while the gray points show the data of the two new GRAVITY observations presented in this work. All three GRAVITY data sets were fitted simultaneously.}
\label{fig:hr8799_cm2_posterior}
\end{figure*}

In Figure \ref{fig:hr8799_cm2_posterior} we show the best-fit spectrum, two- and one-dimensional marginalized posterior, and $P$-$T$ uncertainty envelopes resulting from retrieving the HR8799e observations with Cloud Model 2. This retrieval is discussed in Section \ref{sect:hr8799_cm2}. The priors used for this retrieval are given in Table \ref{tab:prior_hr8799_cm2}.

\begin{table}[t!]
\centering
{ \footnotesize
\begin{tabular}{ll|ll}
 \hline \hline
Parameter & Prior & Parameter & Prior \\ \hline
$T_1$ & $\mathcal{U}(0 , T_2)$ & ${\rm log}(\kappa_0)$ & $\mathcal{U}(-8, 7)$ \\
$T_2$ & $\mathcal{U}(0 , T_3)$ & $\xi$ & $\mathcal{U}(-6,1)$ \\
$T_3$ & $\mathcal{U}(0 , T_{\rm connect})^{\rm (a)}$ & ${\rm log}(P_{\rm base})$ & $\mathcal{U}(-6, 3)$ \& Eq. \ref{equ:pbase} \\
${\rm log}(\delta)$ & $P_{\rm phot} \in [10^{-3}, 100]^{\rm (b)}$ & $\omega$ & $\mathcal{U}(0, 1)$ \\
$\alpha$ & $\mathcal{U}(1,2)$ & $f_{\rm sed}$ & $\mathcal{U}(0, 10)$ \\
$T_0$ & $\mathcal{U}(300 , 2300 )$ & $R_{\rm P}$ & $\mathcal{U}(0.9, 2)$ \\
$\rm C/O$ & $\mathcal{U}(0.1,1.6)$ & ${\rm log}(g)$ & $\mathcal{U}(2, 5.5)$ \\
$\rm [Fe/H]$ & $\mathcal{U}(-1.5,1.5)$ & ${\rm log}(P_{\rm quench})$ & $\mathcal{U}(-6, 3)$ \\ \hline
$f_{\rm SPHERE}$ & $\mathcal{U}(0.8,1.2)$ & $f_{\rm GPI}$ & $\mathcal{U}(0.8,1.2)$ \\ \hline
\end{tabular}
}
\caption{Priors of the non-nominal HR8799e retrieval with Cloud Model 2. $\mathcal{U}$ stands for a uniform distribution, with the two parameters being the range boundaries. The units for the parameters are the same as the ones used for Table \ref{tab:ret_input}. $f_{\rm SPHERE}$ and $f_{\rm GPI}$ are the scaling factors retrieved for the SPHERE and GPI data, respectively. (a) and (b): please see Section \ref{sect:ret_model_setup} for a definition of $P_{\rm phot}$ and $T_{\rm connect}$.}
\label{tab:prior_hr8799_cm2}
\end{table}

\section{Setup of the ANDES disk model}
\label{sect:andes_setup}
Here we describe the setup of the ANDES disk model \citep{2013ApJ...766....8A,2017ApJ...849..130M}, used to study the impact of sophisticated disk chemical modeling on the inferences made about the formation location of planet HR~8799e, see Section \ref{sect:formation}. The gas surface density distribution was described by a tapered power law with an exponent $\gamma = 1$ and a characteristic radius $R_c = 100$~au, the total gas mass in the disk was 0.1~$\rm M_\odot$. The disk midplane temperature was calculated from the stellar and accretion luminosities. The thermal structure of the disk upper layers was calculated using ray tracing in the UV-optical wavelengths. The vertical disk density structure was derived from the hydrostatic equilibrium and was iteratively made consistent with the temperature structure. The gas and dust temperatures were assumed to be equal. The chemical model is based on the gas-grain code ALCHEMIC \citep{2011ApJS..196...25S} with desorption energies updated according to \citet{2017SSRv..212....1C}. An MRN-like power law size distribution \rch{(that is, following \citealt*{mrnref})} with the maximum dust size of 25~micron was used. The average dust radius of 0.37~micron and a dust-to-gas ratio of 0.01 were used for chemical simulations. The assumed stellar parameters representative of HR 8799 were defined using an evolutionary track model for 1.47 Msun star \citep{2008ASPC..387..189Y}. Accretion on the star was described by adding an accretion region with a temperature of 15000 K, contributing to accretion luminosity and UV radiation field. The assumed accretion rate was $10^{-8}$ $\rm M_\odot$/yr. Cosmic rays, X-rays and radioactive nuclides were included as additional sources of ionization. The chemical evolution of the disk was run for 1 Myr, starting from a pre-calculated composition of a 1 Myr old molecular cloud with 
``low metal'' initial abundances  \citep{1998A&A...334.1047L}.

\end{document}